\documentstyle[aps,pre,preprint,amsfonts]{revtex}
\begin{document}

\draft

\title{The Metastate Approach to Thermodynamic Chaos}  
\author{C.M. Newman}
\address{Courant Institute of Mathematical Sciences,
New York University, New York, NY 10012}
\author{D.L. Stein}
\address{Department of Physics, University of Arizona,
Tucson, AZ 85721}

\maketitle

\begin{abstract}
In realistic disordered systems, such as the Edwards-Anderson (EA)
spin glass, no order parameter, such as the Parisi overlap distribution,
can be both translation-invariant and non-self-averaging. The
standard mean-field picture of the EA spin glass phase
can therefore not be valid in any
dimension and at any temperature.  Further analysis 
shows that, in general, when systems have many
competing (pure) thermodynamic states, a single 
state which is a mixture of many of them (as in the
standard mean-field picture)
contains insufficient information to reveal the full thermodynamic
structure.  We propose a different approach, in which
an appropriate thermodynamic 
description of such a system is
instead based on a {\it metastate\/}, which is 
an {\it ensemble\/} of (possibly mixed) thermodynamic states.  This
approach, modelled on chaotic dynamical systems, is  
needed when chaotic size dependence (of finite
volume correlations) is present.  Here replicas 
arise in a natural way, when a metastate
is specified by its (meta)correlations.  The
metastate approach explains,
connects, and unifies such concepts as replica symmetry breaking,
chaotic size dependence and replica
non-independence.  Furthermore, it 
replaces the older idea of non-self-averaging
as dependence on the bulk couplings with the concept of dependence
on the state within the metastate at {\it fixed\/} coupling realization.
We use these ideas to classify possible metastates for the EA model,
and discuss two scenarios introduced by us earlier --- a nonstandard
mean-field picture and a picture intermediate between that
and the usual scaling/droplet picture. 
\end{abstract}



\section{Introduction}
\label{sec:intro}

The nature of the spin glass phase remains a fundamental
and unsolved problem in both condensed matter physics
and statistical mechanics despite over twenty years of 
intensive research.  At a more basic level,
the proper theoretical treatment of systems with quenched disorder
and frustration remains open.  Newer experiments
exhibiting intriguing properties such as aging have not
helped to resolve matters, but have instead intensified
the ongoing debate \cite{Ryan92}.

Spin glasses can be metallic or insulating,
uniaxial or isotropic, mostly crystalline or completely amorphous;
in general they are not confined to a single set
of materials.  The microscopic interactions which give
rise to spin glass behavior may differ considerably
from one material to another.  (For a more extensive discussion,
see the review article by Binder and Young \cite{BY}.)
Nevertheless, in 1975 Edwards and Anderson (EA) \cite{EA} proposed
a simple (and unifying) Hamiltonian to describe the
thermodynamic, magnetic, and dynamical properties of 
realistic spin glasses.  Their basic assumption
was that the essence of spin glass behavior arose
from a competition between quenched ferromagnetic
and antiferromagnetic interactions, randomly
distributed throughout the system.  

While the EA model and its mean-field version,
the Sherrington-Kirkpatrick (SK) model \cite{SK},
remain the primary focus of
theoretical treatments of spin glasses, a number of other
models have also been proposed \cite{BY}.
It is not our aim in this paper to compare the suitability
of these models for the description of all,
or some subset of, laboratory spin glasses.
Here we are concerned instead
with presenting the correct thermodynamic approach to
understanding macroscopic properties of not
only spin glasses, but 
more generally, systems which may have many competing pure states. 
Throughout this paper 
we will often apply our ideas and methods to the EA spin glass
model --- in particular, in its Ising form ---
but our scope is more general and is not confined
to a particular model or a single condensed matter system.
We will begin, however, by considering some of the
very basic open questions which arise in connection with 
the EA Ising spin glass and related models.

These open problems cover both thermodynamic and
dynamical questions.  It is somewhat discouraging that
they persist at such a basic level.
Very slow equilibration times make the analysis of
both laboratory experiments and numerical simulations
difficult; and techniques for the theoretical analysis
of systems with quenched disorder and frustration remain primitive.
So, for example, even though there has been a steady
accumulation of evidence that there exists a
true thermodynamic phase transition in the
EA model (and in real spin glasses), it is fair to say
that the issue is not yet closed (and from the
standpoint of a mathematical proof, or even a convincing 
heuristic argument, remains wide open).  If an
equilibrium phase transition does exist, the lower critical
dimension --- and in particular, whether it is
above or below three --- is 
similarly unknown (see, for example, Refs.~\cite{MEF,MPR,TH}).

Assuming the existence of a phase
transition in some dimension, other 
open thermodynamic issues include the
effect of a magnetic field on the transition;
the number of pure states below the transition;
the correct description of broken symmetry and the
nature of the order parameter; critical properties
at the transition; the role of
quenched disorder and/or frustration, separately
and together, in determining
ground state structure and multiplicity \cite{NS94,NS96a}; 
and the relationship between the properties of mean-field models and realistic
spin glasses.  These are only a subset of such very 
basic questions, which remain the subject of intense debate.  

Dynamical properties are central to spin glass
physics --- here, open problems include
the origin of long relaxation times, the understanding
of frequency-dependent susceptibility experiments,
the origin and interpretation of aging, and 
the nature of metastable states.  As before, this
is only a small sample of outstanding questions.
Tying together both the thermodynamic
and dynamical problems is the general issue
of the nature of broken ergodicity (BE) in spin glasses \cite{Palmer,SA,NS95}.
BE may also serve as a bridge toward investigation
of the relationships, if any, between spin glasses 
and other disordered systems --
structural glasses, electric dipole glasses, 
quadrupolar glasses, and so on \cite{BY}. 

Some thermodynamic questions are of direct experimental relevance:
low-temperature properties cannot be understood
without knowing the nature of low-energy excitations
above the ground state(s); a knowledge
of the critical behavior is required before
properties near $T_c$ (if it exists) can be explained.
It should also be emphasized that many (though not all)
important dynamical questions cannot be properly
understood, or in some cases even posed, without
a correct thermodynamic theory of spin glasses.
For example, what is the relationship, if any, between
the metastable states of a spin glass and the
pure thermodynamic states \cite{Orbach1,Orbach2}?  Moreover,
many experiments, e.g. aging, have been
explained using conflicting theoretical 
pictures \cite{aging1,aging2,aging3,aging4,aging5,aging6,aging7,aging8,aging9,aging10}.
In the absence of conclusive experimental
(or even numerical) data deciding the matter, 
how does one decide among different theories
of the spin glass state, much less explain
the experimental observations?

Our concern in this paper is therefore with the thermodynamic 
nature of spin glasses.  In two recent papers 
\cite{NS96b,NS96c}, we introduced several concepts that we believe are crucial
for providing a correct and complete description of the 
equilibrium statistical mechanics
of spin glasses and other disordered and/or inhomogeneous systems.  Our approach,
modelled on chaotic dynamical systems, is necessary in particular
for understanding systems with competing thermodynamic states.
The unifying idea is that of the metastate \cite{NS96c,AW}, which enables us
to explain and relate chaotic size dependence \cite{NS92}, replica
symmetry breaking \cite{MPV}, replica non-independence,
and overlap (non-)self-averaging.

Using the notion of the metastate, we have classified
allowable thermodynamic ``solutions''
of the spin glass phase (some of which are new), and ruled out
others, including one which has
long dominated the theoretical literature.
In this paper, we will expand and clarify
the ideas presented in Refs.~\cite{NS96b} and \cite{NS96c}, and use
them to present a coherent approach to
the thermodynamics of spin glasses and,
more generally, to disordered and other systems
with many competing states.
We will begin
by discussing a long-standing controversy
over the thermodynamic nature of the spin glass phase.

This controversy focuses on the multiplicity and ordering
of pure states in realistic spin glasses in finite
dimensions, at temperatures below some $T_c>0$ (which, supported
by various arguments, is assumed to exist).  One approach,
which has dominated the spin glass literature for over
a decade, assumes that the main features of Parisi's
solution \cite{Parisi1,Parisi2,Houghton83,MPSTV} of the infinite-ranged
SK model  --- an infinite number of pure states, organized by
an ultrametric overlap structure \cite{MPSTV},
and whose pairwise overlaps are non-self-averaging even in
the thermodynamic limit \cite{MPSTV} --- apply also to realistic
spin glasses.  In this scenario, the number of order
parameters is infinite --- i.e., the order
parameter is a distribution that is a
function of a continuous variable,
and this distribution has a 
characteristic structure, both for a single
realization of the couplings, and for the average
over all such realizations.  The nature of the symmetry
breaking here differs from more conventional kinds,
familiar from studies of various nondisordered systems:  
in Parisi's solution, the spin glass phase(s) exhibit
{\it spontaneously broken replica symmetry\/} of
a nontrivial kind. 

An alternative point of view arises from a
scaling {\it ansatz\/} due to MacMillan \cite{Mac},
Bray and Moore \cite{BM}, and Fisher and Huse \cite{FH86,FH88,HF87a,HF87b}.
This gave rise to a thermodynamic picture 
very different from that implied by the Parisi solution
(although some features, such as chaotic 
dependence of correlation functions on
temperature, are similar in the two pictures).
In particular, the droplet analysis of 
Fisher and Huse \cite{HF87a,HF87b}
led to the conclusion that there exists, at any temperature
and in any finite dimension, at most a pair of pure
states.  (See, however, \cite{E1990} for 
a critique of this prediction.)  Here the order parameter
and the nature of symmetry breaking is markedly
different from that of the Parisi picture.

These two pictures reach opposite conclusions on a number
of other thermodynamic issues; for example, any
external magnetic field destroys the phase transition
in the droplet picture, while that based on
the Parisi solution displays an Almeida-Thouless
line \cite{Alm78}.  (For discussions on
whether numerical evidence supports such a transition,
see Refs.~\cite{Caracciolo,FHcrit}.)  

Both pictures also imply certain
dynamical behavior for spin glasses.  However, although
the physical origins behind various dynamical mechanisms
differ markedly in the two pictures, their observable
consequences are often similar (see, for example, the 
experimental and theoretical discussions on aging in
Refs.~\cite{aging1,aging2,aging3,aging4,aging5,aging6,aging7,aging8,aging9,aging10}),
and most experiments have so far been unable to
distinguish between the two pictures.  (One possible
exception, however, is the set of experiments on 
noise in mesoscopic spin glass samples by Weissman
and collaborators \cite{Weissman}.)

In addition to these pictures, there also exist 
scaling approaches which predict
many pure state pairs at low temperature above
three dimensions \cite{BF1986}.  A number
of other speculative pictures of the spin glass state
have also appeared (see \cite{BY} for a 
more thorough presentation), but it is
probably fair to say that the scaling/droplet and Parisi pictures 
presented above have dominated the discussion
of the nature of the spin glass phase(s).  

The droplet picture of Fisher and Huse
makes a number of clear predictions,
and is relatively easy to interpret for 
realistic spin glasses.  This has not
generally been the case for the Parisi
{\it ansatz\/}, and indeed an important
issue --- although not always recognized
as such --- is to interpret the implications of the
Parisi solution, both thermodynamically and dynamically, for the
spin glass phase in finite dimensions.  A large literature
(see below) exists on the subject, and
as a result a reasonably clear consensus has 
emerged on the thermodynamic structure
of short-ranged spin glasses given that
the Parisi {\it ansatz\/} holds for them also.
We have called this the ``standard SK picture''
in \cite{NS96b} and \cite{NS96c}, and will use that terminology
also throughout this paper.

The main result of \cite{NS96b} was to prove, however, that the
standard SK picture cannot apply to realistic
spin glasses in any dimension and at any temperature.
This result then led in \cite{NS96c} to an observation which is
central to understanding any system with many competing
thermodynamic states:  one should not focus
on any particular (mixed) thermodynamic state, which cannot
provide sufficient information to describe the thermodynamic
structure; instead, one must consider the {\it metastate\/},
which is essentially a probability distribution over
the thermodynamic states.  One important consequence
of \cite{NS96b} and \cite{NS96c} is to redefine the meaning of non-self-averaging,
and to show that most quantities of interest can be defined
for a {\it single\/} realization of the disorder (including
those which had been thought to be non-self-averaging
in the thermodynamic limit).
One can then focus on, and make meaningful
statements about, a particular sample rather than 
an ensemble of samples.  This feature should hold also for
nondisordered (e.g., inhomogeneous) systems in general.

Using the metastate approach, we were able
to narrow down the possible thermodynamic
structures for realistic spin glasses.  
One of these is the scaling/droplet
picture; some are new.  Finally, we proposed
a possible picture which incorporates some
of the features of the Parisi solution for
the SK model.  In fact, this is the 
``maximal'' mean-field picture allowable
for realistic spin glasses, but it differs
considerably from the familiar standard
SK picture presented in the literature.
We call this new scenario the ``nonstandard SK picture''
and will discuss it in Section \ref{sec:nonstandard}.
One important lesson from our analysis is that,
for disordered systems, the features which
characterize the system in very large but
finite volumes may lead to a misleading thermodynamic
picture if straightforwardly extrapolated to infinite 
volumes.  This is of potential importance, 
for example, in interpreting numerical results.  There are previously
unsuspected intricacies involved in taking
the thermodynamic limit for certain disordered systems.

The plan of the paper is as follows.  In Section~\ref{sec:EA} we
review some basic features of the EA model and discuss its finite-
and infinite-volume Gibbs states.  We discuss the problem
of whether many pure states may exist at some dimension and
temperature, and show that the answer is independent
of coupling realization.
In Section~\ref{sec:SK} we introduce the SK model
and the Parisi {\it ansatz\/} for its thermodynamic
structure.  In Section~\ref{sec:standard} we discuss
the standard mean-field picture for realistic
spin glass models in finite dimension.  We then
show that this picture cannot hold in any dimension
and at any temperature.  We also provide an 
explicit construction of a non-self-averaged
thermodynamic state whose overlap
distribution function must be self-averaged.
In Section~\ref{sec:meta} we describe a new
approach to the thermodynamics of systems
with many competing states, based on
the idea of the {\it metastate\/}, an
{\it ensemble\/} of thermodynamic states.
We also present some of the possible scenarios for
the metastate of the EA model, including one
that is intermediate between the scaling/droplet
and mean-field pictures.
In Section~\ref{sec:rsbetc} we show
how replicas arise naturally within this
approach, and how the older idea of replica
symmetry breaking is understood and unified with
newer ideas of dispersal of the metastate
and replica non-independence.  A replacement
for the usual definition of non-self-averaging
is also presented.  In Section~\ref{sec:nonstandard}
we introduce the maximal mean-field picture allowable
for realistic spin glasses, and show that its thermodynamic
features are considerably different from those of
the more familiar picture (which cannot hold).
Finally, in Section~\ref{sec:conclusions}, we
summarize our main results and discuss their implications for the study
of spin glasses and, in a larger framework, 
systems with many competing states in general.

\section{The Edwards-Anderson model}
\label{sec:EA}

The Edwards-Anderson (EA) model \cite{EA} on a cubic lattice
in $d$ dimensions is described by the Hamiltonian
\begin{equation} 
\label{eq:EA}
{\cal H}_{\cal J}(\sigma)= -\sum_{\langle x,y\rangle} J_{xy} \sigma_x \sigma_y\quad ,
\end{equation}
where ${\cal J}$ denotes the set of couplings $J_{xy}$ and where
the brackets indicate that the sum is over nearest-neighbor
pairs only, with the sites $x,y\in Z^d$.  We will take the spins
$\sigma_{x}$ to be Ising, i.e., $\sigma_{x}=\pm 1$; 
although this will affect the details of
our discussion, it is unimportant for our main conclusions.
The couplings $J_{xy}$ are quenched, independent, identically
distributed random variables; throughout the paper we will
assume their common distribution to be symmetric about zero
(and usually with the variance fixed to be one).  
The most common examples are the
Gaussian and $\pm J$ distributions.  

Eq.~(\ref{eq:EA}) is the EA Ising Hamiltonian for
an infinite-volume spin glass on $Z^d$; it is
important also to define the EA model on a finite
volume, given specified boundary conditions.
Let $\Lambda_L$ be a cube of side $2L+1$
centered at the origin; i.e., 
$\Lambda_L= \{-L, -L+1, \cdots ,L\}^d$.
The finite-volume EA Hamiltonian is then
just that of Eq.~(\ref{eq:EA}) confined
to the volume $\Lambda_L$, with the spins
on the boundary $\partial\Lambda_L$ of
the cube obeying the specified boundary condition.
(The boundary $\partial\Lambda_L$ of the volume $\Lambda_L$ consists 
of all sites not in $\Lambda_L$ with one nearest
neighbor belonging to $\Lambda_L$.)
For example, the Hamiltonian with {\it free\/}
boundary conditions is simply
\begin{equation} 
\label{eq:free}
{\cal H}_{{\cal J},L}^{f}(\sigma)= -\sum_{\langle x,y\rangle\in\Lambda_L} J_{xy} \sigma_x \sigma_y\quad .
\end{equation}
Another important boundary condition, called a fixed b.c., is where
the value of each spin on the boundary is specified.  If we denote
by $\overline\sigma$ the specified boundary spins, 
then the Hamiltonian becomes
\begin{equation} 
\label{eq:fixed}
{\cal H}_{{\cal J},L}^{\overline\sigma}(\sigma)= 
{\cal H}_{{\cal J},L}^{f}(\sigma)
-\sum_{\stackrel{\langle x,y\rangle}{x\in\Lambda_L,y\in\partial\Lambda_L}} 
J_{xy} \sigma_x \overline\sigma_y\quad .
\end{equation}

We will frequently employ a familiar and commonly used boundary
condition, namely periodic boundary conditions, where each face
of the cube $\Lambda_L$ is identified with its opposite face.  These
are generally thought of as minimizing the effects of the boundary
(but see van Enter \cite{E1990}), and allow us to construct
manifestly translation-covariant states.  

Given the EA Hamiltonian ${\cal H}_{{\cal J},L}$ 
on a finite volume $\Lambda_L$ with a specified
boundary condition (e.g., free or fixed or periodic,
but without the boundary condition superscript
here), we can now define the finite-volume
Gibbs distribution $\rho_{{\cal J},\beta}^{(L)}$ on $\Lambda_L$ at
inverse temperature $\beta$:
\begin{equation}
\label{eq:finite}
\rho_{{\cal J},\beta}^{(L)}(\sigma)=Z_{L,\beta}^{-1} \exp \{-\beta
{\cal H}_{{\cal J},L}(\sigma)\}\quad ,
\end{equation}
where the partition function $Z_{L,\beta}$ is such that the sum
of $\rho_{{\cal J},\beta}^{(L)}$ over all spin configurations in $\Lambda_L$
yields one. In addition to the boundary conditions mentioned so far,
one also considers so-called mixed boundary conditions 
where the Gibbs distribution
is a convex combination of the fixed boundary condition distributions
for a given $L$ with the weights for the different 
$\overline\sigma$'s adding up to one.

$\rho_{{\cal J},\beta}^{(L)}(\sigma)$ is a finite-volume probability
measure, describing at fixed $\beta$ the likelihood of appearance, 
within the volume $\Lambda_L$, of a given
spin configuration  obeying the specified boundary
condition.  Equivalently, the measure is
specified by the set of all correlation functions 
within $\Lambda_L$,
i.e., by the set of all
$\langle\sigma_{x_1}\cdots\sigma_{x_m}\rangle$ for arbitrary
$m$ and arbitrary $x_1,\ldots,x_m\in\Lambda_L$.

{\it Thermodynamic\/} states are described by {\it infinite\/}-volume
Gibbs measures, and therefore can be thought of (and indeed, constructed) as
a limiting measure of a sequence, as $L\to\infty$, of such finite-volume measures (each with
a specified boundary condition, which may remain the same or may change
with $L$) \cite{NoteDLR}.  
The idea of a limiting measure
can be made precise by requiring that every $m$-spin correlation function,
for $m=1,2,\ldots$, possesses a limit as $L\to\infty$.

It is clear that, if there is more than one thermodynamic state
(at a given temperature) and if arbitrary boundary
conditions are allowed for each $L$, different
sequences (of volumes and/or boundary conditions) can have different limiting measures.
What is less obvious, but will have important consequences for
spin glasses, is that if many thermodynamic states exist, 
a sequence of measures each having the {\it same\/} (e.g., periodic or free)
boundary condition may not even {\it have\/} a limit \cite{NS92}.
This phenomenon, which we call {\it chaotic size dependence\/}, will
be more fully described in Section~\ref{sec:meta}.  Because of
compactness (i.e., because each of the correlations determining
the measure takes values in $[-1,1]$, a bounded closed interval),
it follows,
however, that every such infinite sequence will have some  
subsequence(s) with a single limit, 
so that we are guaranteed the existence of at least one
thermodynamic state (i.e., infinite-volume Gibbs distribution).
At sufficiently high temperatures it is rigorously known (see below)
that there exists only one such state (limiting Gibbs measure), which of course is the
paramagnetic state.  If the spin-flip symmetry present in the EA
Hamiltonian Eq.~(\ref{eq:EA}) is spontaneously
broken above some dimension $d_0$ and below some
temperature $T_c(d)$, there will be at least a {\it pair\/} of
limiting measures, such that their even-spin correlation
functions will be identical, and their odd-spin correlation
functions will have the opposite sign.  Assuming that
such broken spin-flip symmetry indeed exists for $d>d_0$
and $T<T_c(d)$, the question
of whether there exists {\it more than one\/} such limiting pair
(of spin-flip related infinite-volume Gibbs distributions) is a central
unresolved issue for the EA and related models.

Thermodynamic states may or may not be mixtures
of other states.  If a Gibbs state $\rho_{{\cal J},\beta}$
can be decomposed according to
\begin{equation}
\label{eq:mixed}
\rho_{{\cal J},\beta}=\lambda\rho^{1}_{{\cal J},\beta}+(1-\lambda)\rho^{2}_{{\cal J},\beta}\quad ,
\end{equation}
where $0 < \lambda < 1$ and $\rho^{1}$ and  $\rho^{2}$ are also
infinite-volume Gibbs states (distinct from $\rho$), 
then we say that $\rho_{{\cal J},\beta}$ is a {\it
mixed\/} thermodynamic state or simply, mixed state.  (A mixed state may,
of course, have many, perhaps infinitely many, states in its
decomposition.)  The meaning of Eq.~(\ref{eq:mixed}) can be understood as
follows:  any correlation function computed using the Gibbs distribution
$\rho_{{\cal J},\beta}$ can be decomposed in the following way:
\begin{equation}
\label{eq:decomposed}
\langle\sigma_{x_1}\cdots\sigma_{x_m}\rangle_{\rho_{{\cal J},\beta}} = 
\lambda\langle\sigma_{x_1}\cdots\sigma_{x_m}\rangle_{\rho^{1}_{{\cal
J},\beta}}
+(1-\lambda)\langle\sigma_{x_1}\cdots\sigma_{x_m}
\rangle_{\rho^{2}_{{\cal J},\beta}}\quad .
\end{equation}

If a state cannot
be written as a convex combination of any other infinite-volume Gibbs
states, it is called a {\it pure\/} (or extremal) state.  As an illustration, the
paramagnetic state is a pure state, as are each of the positive and negative
magnetization states in the Ising ferromagnet.  In that same system, the Gibbs
state produced by a sequence of increasing
volumes, at $T<T_c$, using only periodic or free boundary conditions is a mixed
state, decomposable into the positive and negative magnetization states,
with $\lambda=1/2$.  A pure state $\rho_P$ 
can be intrinsically characterized
by a {\it clustering property\/} (see, e.g., \cite{Georgii,vEvH}),
which implies that for any fixed $x$,
\begin{equation}
\label{eq:clustering}
\langle\sigma_x\sigma_y\rangle_{\rho_P} - 
\langle\sigma_x\rangle_{\rho_P}\langle\sigma_y\rangle_{\rho_P}\ \to 0 ,
\qquad \vert y \vert\to\infty\quad ,
\end{equation}
and similar clustering for higher order correlations.

Let $\eta({\cal J},d,\beta)$ now denote the number of pure states in the EA
model for a specific coupling realization ${\cal J}$.  For any $d$ and ${\cal J}$ 
this equals one at sufficiently low $\beta$ (except
for a set of ${\cal J}$'s with zero probability according to the
underlying disorder distribution---see, e.g., Chapter 3
of \cite{Zurich} and the references cited there).  Recall that the droplet
picture predicts that $\eta({\cal J},d,\beta)\le 2$ for all $d$ 
and $\beta$, while
the SK picture assumes that $\eta({\cal J},d,\beta)= \infty$ for 
$d$ above some (unknown) $d_0$ and $\beta>\beta_c(d)$. 

A reasonable question might then be, could the answer (at fixed $\beta$ and $d$)
depend on ${\cal J}$?  What if $\eta=2$ for half the coupling realizations
(i.e., for a set of coupling realizations with probability $1/2$),
and infinity for the other half?  As it turns out, this cannot happen:  
for a fixed coupling distribution, $\eta$ at some $(d,\beta)$ must have the same value
for {\it all\/} instances ${\cal J}$ chosen from the disorder
distribution (or more precisely, for {\it almost every\/} ${\cal J}$ ---
i.e., except for a set with 
probability zero).
In other words, $\eta({\cal J},d,\beta)$ is {\it self-averaged\/};
for fixed $d$ and $\beta$, it is a constant almost surely 
as a function of ${\cal J}$.

The above statement is mathematically rigorous, but since its
proof, and that of all other theorems which appear in this
paper, have appeared elsewhere (see, e.g., \cite{NSBerlin,Zurich}),
we here recount only the central arguments.
(These arguments will be useful later 
when we discuss possible scenarios for the
thermodynamic structure of the spin glass phase in the EA model.)
We first note that $\eta({\cal J},d,\beta)$ is clearly
translation-invariant;  i.e., if all couplings are 
translated by any lattice vector $a$, so that each $J\to J^a$
(i.e., $J_{xy}\to J_{x+a,y+a}$), 
the function is unchanged:  $\eta({\cal J},d,\beta)=\eta({\cal
J}^a,d,\beta)$.  We next note that the 
disorder distribution $\nu({\cal J})$
(e.g., an independent Gaussian distribution of mean zero and variance one
at each bond on the lattice) is both translation-invariant (trivially) and
{\it translation-ergodic\/}.  Translation-ergodicity means that 
for $\nu$-almost every ${\cal J}$,
the (spatial) average of translates $\hat f({\cal J}^a)$ of any
(measurable) function $\hat f$ on ${\cal J}$ equals 
the $\nu$-average of $\hat f$.  (As a trivial example,
consider a $1d$ problem where 
the function $\hat f ({\cal J})$ is just the coupling value $J_{01}$ at 
a given location on the line.
The average of $\hat f ({\cal J}^a)$ along the line is clearly $0$; so is
the distribution average at any site.  Similarly, for 
the function $\hat f ({\cal J}) = (J_{01})^k$,
the spatial average along the line equals the distribution average at a site,
i.e., the $k$th moment of the random variable $J_{01}$.)
That translation-ergodicity in several dimensions holds
analogously to the more familiar one-dimensional case
seems to have first been shown by Wiener \cite{Wiener1939}.

The assumption that $\hat f$ be measurable is a necessary,
but somewhat technical requirement.
A proof that $\eta({\cal J},d,\beta)$ satisfies the necessary 
measurability properties appears in \cite{NSBerlin,Zurich}, and
will not be discussed further here, except to note that
measurability of a function $\hat f$ is the minimal requirement
for having a well defined meaning for the average of $\hat f$
over the disorder distribution $\nu$.

Returning to the main argument, we note that because $\eta({\cal J},d,\beta)$
is a translation-invariant function of random variables  ${\cal J}$
whose distribution is translation-ergodic, by averaging
$\eta({\cal J}^a,d,\beta) = \eta({\cal J},d,\beta)$ over translates,
it follows that $\eta({\cal J},d,\beta)$ equals a constant $\eta(d,\beta)$
almost surely (i.e., for almost every ${\cal J}$).  
$\eta(d,\beta)$ is the distribution average
of $\eta({\cal J},d,\beta)$ and it could of course depend on the {\it distribution\/}
from which the couplings are chosen, but not on any specific realization
chosen from a {\it fixed\/} distribution.  

The same line of reasoning used here to show that the number of pure states
at fixed $d$ and $\beta$ is the same for almost every realization ${\cal J}$
was used in \cite{NS96b} to rule out the standard SK picture.  This will be discussed later
in Section~\ref{sec:standard}; but first we present a discussion of
the infinite-ranged SK model and the Parisi solution.

\section{Mean-field theory and the Parisi solution}
\label{sec:SK}

The Sherrington-Kirkpatrick (SK) model has played an important role in spin
glass physics for several reasons. First among these is that
it is one of the few (nontrivial) spin glass models which 
is (generally) believed
to have been solved.  Moreover, the proposed solution, due to Parisi
\cite{Parisi1,Parisi2,MPSTV} admits 
a striking type of symmetry breaking, called
{\it replica symmetry breaking\/} (RSB), of a form previously
unseen in other (nondisordered) systems.  The possibility that RSB 
plays an important role in the physics of 
{\it realistic\/} (i.e., finite dimensional) 
spin glasses and, possibly, other complex systems has
generated a substantial literature (see, for example,
Refs.~\cite{MPR,Orbach1,Orbach2,MPSTV,MP,BMY1,BMY2,Parisi3,VHO,BCPR,FPV,Ritort94,LD,MPRR})
and remains controversial.

The SK model is simply the infinite-ranged version of the EA model  
and thus has no spatial structure. 
The volume $L^d$ is replaced by $N$, the number of spins, and
the Hamiltonian is
\begin{equation} 
\label{eq:SK}
{\cal H}_{{\cal J},N}(\sigma)= -{1\over\sqrt{N}}\sum_{i>j=1}^N J_{ij} \sigma_i \sigma_j\quad ,
\end{equation}
where the factor $1/\sqrt{N}$ ensures that the energy per spin remains
finite (and nonzero) as
$N\to\infty$ (given that, as before, the distribution of each $J_{ij}$ is
symmetric about zero and has variance one).  Because there is no natural
sense of a boundary in this model, one usually considers simply a sequence
of Hamiltonians of the form (\ref{eq:SK}) with increasing $N$.  The
probability measure on spin configurations in this model is given by
\begin{equation}
\label{eq:infinite}
\rho_{{\cal J},\beta}^{(N)}(\sigma)=Z_{N,\beta}^{-1} \exp \{-\beta
{\cal H}_{\cal J,N}(\sigma)\}\quad .
\end{equation}

It has been known for many years \cite{Brout}
that a correct treatment of quenched disorder involves an averaging
(over the couplings) of the 
free energy and other extensive variables rather
than of the partition function.  The replica trick
\cite{EA,Kac,Edwards} was introduced as a tool for 
doing such an average; because of the lack of spatial structure
in the SK model, it is especially well suited for this approach.
Using the replica trick,
SK demonstrated the existence of a phase
transition, but found that the resulting low-temperature phase was
unphysical \cite{SK}.
It is currently believed that their solution was unstable
because it was replica-symmetric.  Several attempts were made to introduce
solutions which broke the replica symmetry \cite{Blandin}, but it
is now thought that the correct procedure to break replica
symmetry in the low-temperature phase of the SK model was 
the one introduced by Parisi \cite{Parisi1}.

The Parisi solution to the SK model is both stable and agrees well with
numerical results \cite{MPV}; moreover, some of its essential features
can be rederived without the use of replicas, primarily through a cavity
method \cite{DT,MPV86}.  Parisi's approach suggests that there are many
pure states of the infinite-ranged model, organized in a highly specific
manner which characterizes the SK spin glass phase and its mode
of symmetry breaking.  Although Parisi's solution
predicts many other important features of the spin glass phase\cite{BY,MPV}, we will focus
here only on its aspects regarding symmetry breaking and order parameters.

We first need to comment on what is meant by ``pure state'' in the SK
model, since a precise definition is not available and its meaning
remains quite unclear.  Other approaches to spin glass mean-field
theory (e.g., the TAP equations \cite{TAP}) had already suggested the
existence of many states at low temperature, in the sense that many solutions could
be found which were extrema of the free energy, some subset of which
were believed to be minima \cite{BM1980}.  It had further been argued
that they were separated by barriers which diverged in the thermodynamic
limit \cite{BM1981,MY1982}.  These are what have come to be called \cite{MPV} the ``pure
states'' of the SK model.  The clustering property described by 
Eq.~(\ref{eq:clustering}) cannot be used in an infinite-ranged model, which
has no spatial structure, but it has been suggested \cite{BY,MPV}
that it can be replaced by
\begin{equation}
\label{eq:SKclustering}
\lim_{N\to\infty}\langle\sigma_i\sigma_j\rangle_{\beta,N} -
\langle\sigma_i\rangle_{\beta,N}
\langle\sigma_j\rangle_{\beta,N}\ = 0 , \qquad (i\ne j)\quad ,
\end{equation}
where averages are taken using the distribution corresponding to one of
these pure states.  The meaning of the averaging in 
Eq.~(\ref{eq:SKclustering}) is poorly
defined, however.  Because the strength of the random couplings
scales to zero as $N\to\infty$, it is unclear what
meaning, if any, can be ascribed to the notion of 
nontrivial thermodynamic
states, pure or mixed.  In the EA model,
on the other hand, methods do exist, as 
will be briefly discussed in the next section,
to construct just such states, for almost every ${\cal J}$.
This contrast between the SK and EA models 
will be seen to be significant.

However, in accordance with the usual practice, we will ignore
these complications in what follows, though
keeping in mind that the meaning of pure state in the SK context remains
vague.  Using a replica analysis, 
Parisi found that the SK spin glass state could
be described properly only with an {\it infinite\/} number
of order parameters, describing the relations among the
many pure states.  This requires the introduction of a new
random variable which describes the {\it replica overlap\/},
\begin{equation}
\label{eq:overlap}
Q_N={1\over N}\sum_{i=1}^N\sigma_i\sigma'_i \; ,
\end{equation}
where the spin configurations $\sigma$ and $\sigma'$
are chosen independently from the distribution $\rho_{{\cal
J},\beta}^{(N)}(\sigma)$
given by Eq.~(\ref{eq:infinite}).
(Technically, $\sigma$ and $\sigma'$ are said to be
chosen from the {\it product distribution\/} $\rho_{{\cal
J},\beta}^{(N)}(\sigma)\rho_{{\cal J},\beta}^{(N)}(\sigma')$).
It is clear that $-1<Q_N<1$ for any $N$.  (The subscript
$\beta$ will be suppressed in expressions related
to replica overlaps and their distributions.  It is understood
that all calculations take place at fixed $\beta$, and
the results depend on $\beta$.)

The role of order parameter in the Parisi theory is
played not by a single variable, but rather by the
probability density $P_{{\cal J},N}(q)$ of the random variable $Q_N$
(or functions closely related to it); i.e., $P_{{\cal J},N}(q)\ dq$
is the probability that the random variable $Q_N$ takes on a value between
$q$ and $q+dq$ (for fixed ${\cal J}$, $N$ and $\beta$).
Above the critical temperature (i.e., in the paramagnetic state), the distribution
of $Q_N$ converges to a delta-function at zero as $N\to\infty$.  Below
this temperature, however, the presumed existence of many states
gives rise to a rich and nontrivial behavior of $P_{{\cal J},N}(q)$.
In particular, Parisi found that in the spin glass phase, 
$P_{{\cal J},N}(q)$ approximates a sum of many delta-functions,
with weights and locations depending on ${\cal J}$ even in the limit
$N\to\infty$.  This is the first appearance of
non-self-averaging (NSA), which plays a central role in the Parisi
theory of the spin glass phase.

The usual explanation given for this behavior is that for
large $N$ the Gibbs measure $\rho_{{\cal J},\beta}^{(N)}$ 
given by Eq.~(\ref{eq:infinite})
(for $\beta>\beta_c$) has a decomposition into many pure states
$\rho_{\cal J}^{\alpha}$, where $\alpha$ indexes the pure states: 
\begin{equation}
\label{eq:sum}
\rho_{{\cal J}}^{(N)}(\sigma)\approx\sum_
\alpha W_{\cal J}^\alpha\rho_{\cal J}^\alpha (\sigma)\ ,
\end{equation}
where $W_{\cal J}^\alpha$ denotes the weight of pure state $\alpha$ and
the dependence on the inverse temperature $\beta$ has been suppressed.
(The use of the approximation sign is necessary because of
the haziness of the meaning of pure state, as discussed above.)
Granted the existence of these pure states, one can then
consider the case where $\sigma$ is drawn from the distribution $\rho_{\cal J}^\alpha$ and $\sigma'$
independently from $\rho_{\cal J}^\gamma$; then 
the expression in Eq.~(\ref{eq:overlap}) equals its thermal mean,
\begin{equation}
\label{eq:qab}
q_{\cal J}^{\alpha\gamma}\approx {1\over N}
\sum_{i=1}^N\langle\sigma_i\rangle^\alpha
\langle\sigma_j\rangle^\gamma \quad .
\end{equation}
Finally, the density $P_{{\cal J},N}(q)$ is then given by
\begin{equation}
\label{eq:PJ(q)}
P_{{\cal J},N}(q)\approx\sum_{\alpha,\gamma}W_{\cal J}^\alpha W_{\cal J}^\gamma
\delta(q-q_{\cal J}^{\alpha\gamma})\quad .
\end{equation}
These expressions can be made precise for the EA model, as will
be seen in the next section.  

The $q_{\cal J}^{\alpha\gamma}$'s also exhibit NSA for arbitrarily
large $N$, except for the trivial cases $\alpha=\gamma$
and $\alpha=-\gamma$ (where the minus sign denotes a global spin flip), 
which correspond respectively to the self-overlaps
$q_{EA}$ and $-q_{EA}$ (with no dependence on
${\cal J}$ or $\alpha$).  Why don't we then simply examine the $N\to\infty$ limit
of the $q_{\cal J}^{\alpha\gamma}$'s and their distribution?  
{\it A~priori\/} it might seem that, even though the states themselves
are not well-defined for infinite $N$, their {\it overlaps\/} might still
have a well-defined limit.  It can be proved, however,
that {\it the existence of the $N\to\infty$ limit 
(where the limit is taken in a ${\cal J}$-independent manner) of the
distribution of the $q_{\cal
J}^{\alpha\gamma}$'s is inconsistent with NSA\/} \cite{NS92}.
This is the first appearance of chaotic size dependence,
which will be seen later to play a central role in the analysis
of systems with many competing states.

Even though the decomposition of Eq.~(\ref{eq:sum}) is presumed to consist
of infinitely many states as $N\to\infty$, it is believed \cite{MPV} that
relatively few of them have non-negligible weight (and are therefore
thermodynamically significant).  These lowest-lying states
are believed to have free energy differences
of order one (for arbitrarily large $N$), and their density 
rises exponentially \cite{DT,MPV86} at the lowest energies.

So far we have only discussed the overlaps among pairs
of pure states.  The relationships among triples
of pure states were also investigated \cite{MPSTV},
and were found to have an ultrametric structure.
That is, the Hamming distances (determined by the overlaps) among
any three pure states are such that the largest two
are always equal (with the third smaller than or
equal to the other two).

The main features of the Parisi analysis of the SK
model, relevant to the ordering of the spin glass
phase, are then the following:

\noindent 1)  The spin glass phase consists of a mixture of
infinitely many
pure states.  Two replicas have non-negligible probability
of appearing in different pure states (not related by a trivial
global spin flip).  This is one interpretation of 
spontaneous replica symmetry breaking (RSB). 

\noindent 2)  For fixed ${\cal J}$, the probability density
$P_{\cal J}(q)$ consists of a sum of (approximate) delta-functions
at discrete locations $q$ such that $-q_{EA}<q<q_{EA}$, 
and a pair always at $\pm q_{EA}$.
The weights and locations of the delta-functions,
excluding the pair at $\pm q_{EA}$,
depend on ${\cal J}$, even as $N\to\infty$ (NSA).

\noindent 3)  Because of this variation with ${\cal J}$ of 
the ``interior'' delta-functions,
the average $P(q)=\overline{P_{\cal J}(q)}$
over all (uncountably many) coupling realizations has a continuous
(and nonzero) component for $q$ between the delta-functions
at $\pm q_{EA}$, for $0<T<T_c$.

\noindent 4)  The locations of the delta-functions
in 2) have an ultrametric structure.

In the next section, we examine the meaning of 
the Parisi picture applied to the EA model.

\section{The standard SK picture}
\label{sec:standard}

If the Parisi~solution of the SK~model describes
the nature of the spin glass phase in realistic spin glasses,
as frequently supposed \cite{MPR,Orbach1,Orbach2,MPSTV,MP,BMY1,BMY2,Parisi3,BCPR,FPV,Ritort94,LD,MPRR},
what should be its thermodynamic properties?  A 
description along these lines has emerged in the literature
(see, for example, Refs.~\cite{MPV,Parisi3,Ritort94,MPRR,RBY}) over
the past decade and a half.  This scenario, which
we have called the ``standard SK picture'' \cite{NS96c}, is
the most straightforward extrapolation
of the main features of the Parisi
solution to infinite volume spin glasses in finite dimension
and is presented in this section as
a precise description of the usual 
presentations in
the literature (see,
for example, Refs.~\cite{BY,MPV,Caracciolo,RBY}).

As discussed in Section~\ref{sec:EA}, the meaning
of pure states in the EA model (and other realistic models)
is clear --- they are extremal infinite volume Gibbs states
(i.e., thermodynamic states which cannot be decomposed
as in Eq.~(\ref{eq:mixed}); equivalently, they satisfy
the clustering property as in Eq.~(\ref{eq:clustering})).
It is natural then to replace the approximate relation
Eq.~(\ref{eq:sum}) with an equality
\begin{equation}
\label{eq:sumea}
\rho_{\cal J}(\sigma)=\sum_\alpha W_{\cal J}^\alpha\rho_{\cal J}^\alpha (\sigma)\ ,
\end{equation}
where $\rho_{\cal J}(\sigma)$ is an infinite volume mixed Gibbs state 
(at fixed temperature) for a particular coupling realization ${\cal J}$, 
and the $\rho_{\cal J}^\alpha$ are
pure states for that ${\cal J}$.  There may be many such mixed states, so we specify
the one above as that produced in some natural way 
(to be specified later) by a
sequence  as $L\to\infty$ of finite volume Gibbs distributions on 
cubes $\Lambda_L$ with boundary
conditions, such as periodic, not depending on the coupling realization. 
Periodic boundary conditions minimize, in some sense, 
the effect of a boundary and are thus 
a natural analogue to the lack of boundary conditions in
the SK model. It should be noted however that there is some
possibility of different behavior for periodic as opposed to, say,
free boundary conditions \cite{E1990}.

We digress momentarily to discuss briefly
two important points.  The first is that, while the
notion of an infinite volume (pure or mixed) state is well defined
for nearest neighbor models, it is less so for
systems with very long-ranged interactions, such as
RKKY.  Our arguments that are based on the homogeneity 
of the disorder, presented below,
will still apply to these systems, but this point should be kept in mind.

The second point is that it is necessary that $\rho_{\cal J}$,
obtained from the natural limit procedure discussed above, be
defined for almost every ${\cal J}$ chosen from the disorder
distribution $\nu$ (and be measurable in its dependence
on ${\cal J}$ --- see
Section~\ref{sec:EA}). 
While this may seem like a technical point of little physical
consequence, it actually plays a crucial role in any thermodynamic
treatment of systems with many competing states \cite{NS96b,NS96c,NS92}.
We will come back to this point in a little while.

Returning to the standard SK picture, we note that the other equations 
in Section~\ref{sec:SK} are similarly replaced
with their exact EA counterparts.  The overlap random variable
becomes
\begin{equation}
\label{eq:overlapea}
Q=\lim_{L\to\infty}|\Lambda_L|^{-1}\sum_{x\in\Lambda_L}\sigma_x\sigma'_x
\end{equation}
where $\sigma$ and $\sigma'$ are chosen, similarly to before, from
the product distribution $\rho_{\cal J}(\sigma)\rho_{\cal J}(\sigma')$. 
If $\sigma$ is drawn from $\rho_{\cal J}^\alpha$ and $\sigma'$ from
$\rho_{\cal J}^\gamma$, then it follows that the overlap is the constant
\begin{equation}
\label{eq:qabea}
q_{\cal J}^{\alpha\gamma} = 
\lim_{L\to\infty}|\Lambda_L|^{-1}\sum_{x\in\Lambda_L}\langle\sigma_x\rangle^\alpha
\langle\sigma_x\rangle^\gamma \quad .
\end{equation}
The probability distribution $P_{\cal J}(q)$ of $Q$ is therefore given by
\begin{equation}
\label{eq:PJ(q)ea}
P_{\cal J}(q)=\sum_{\alpha,\gamma}W_{\cal J}^\alpha W_{\cal J}^\gamma
\delta(q-q_{\cal J}^{\alpha\gamma})\quad .
\end{equation}

So this SK picture for the EA model (or realistic
spin glasses in general) includes the same four
features presented at the end of the previous section
(except the word ``approximate'' in the second of these
should be deleted);
their meanings are now precise.  There are other
elements of the standard SK picture --- e.g., energy
gaps of order one separating the lowest-lying states
in any large volume and an exponentially increasing density
at the lowest energies --- but these will
not play a central role in what follows.

We turn now to the question
of whether the standard SK picture can be valid
in any dimension or at any temperature.  This question has two parts. 

First, does there exist some
natural construction which begins with the finite volume Gibbs states
$\rho_{{\cal J},\beta}^{(L)}(\sigma)$ of Eq.~(\ref{eq:finite}),
takes $L\to\infty$, and ends with a (non-self-averaged) infinite volume
state, $\rho_{\cal J,\beta}$ (possibly mixed), 
and its accompanying overlap distribution 
$P_{\cal J}(q)$?  By ``natural'' we mean not only the usual sense
of the term but also that
the construction result in a thermodynamic state $\rho_{\cal J}$
for almost every ${\cal J}$. In particular, we want the limit procedure
(e.g., choice of boundary conditions or sequence of cube sizes) 
to be chosen independently of any specific ${\cal J}$.  This will help guarantee that
the ${\cal J}$-dependence generated by this procedure is
measurable, and therefore that averages (e.g., of the moments of $Q$)
can be taken with respect to the disorder distribution.
We also emphasize that we are interested only in
procedures which result in non-self-averaged infinite-volume
states (i.e., at least some correlation functions computed within
such a state depend on ${\cal J}$).  Recall that for the SK model, the
very notion of such a ${\cal J}$-dependent infinite-volume state is unclear.

Second, can such a $P_{\cal J}(q)$ exhibit all the
essential features of the SK picture, including those 
described by the four features
above?  

The answers to these two parts, given in \cite{NS96b} are, respectively,
yes and no. We will explain our construction of $\rho_{{\cal J}}$ (which is
somewhat technical) and thus the ``yes'' answer to part one later 
in this section. Meanwhile, we mention one
crucial feature of the resulting $\rho_{{\cal J}}$, which plays a key role
in the ``no'' to part two. That feature is translation-covariance;
i.e., under the translation of
${\cal J}$ to ${\cal J}^a$, where $J^a_{xy}=
J_{x+a,y+a}$ for each $J_{xy}$, $\rho_{\cal J}$ transforms so that
\begin{equation}
\label{covariant}
\rho_{{\cal J}^a}(\sigma_{x_1}=\sigma_1,\dots,\sigma_{x_m}=\sigma_m)=
\rho_{\cal J}(\sigma_{x_1-a}=\sigma_1,\dots,\sigma_{x_m-a}=\sigma_m).
\end{equation}
The conceptual significance of translation-covariance is that the mapping
from ${\cal J}$ to $\rho_{\cal J}$, being a natural one, should not (and in
our construction does not) depend on the choice of an origin.  It also
implies the technical conclusion \cite{NSBerlin,Zurich} that this procedure
leads to a limiting infinite-volume overlap distribution function
$P_{\cal J}(q)$, which exists for almost every ${\cal J}$ and depends measurably
on ${\cal J}$ (guaranteeing that averages of $q$-dependent functions can
be taken over the couplings).
To begin our
answer to the second part of the question, we see what the
translation-covariance of $\rho_{\cal J}$ implies about $P_{\cal J}$.

By translation-covariance of $\rho_{\cal J}$, the overlap random variable
$Q_{{\cal J}^a}$ has the same distribution as the random variable
$Q_{\cal J}^{-a}$, where 
\begin{equation}
\label{invariance}
Q^{-a}_{\cal J}\equiv \lim_{L\to\infty}|\Lambda_L|^{-1}\sum_{x\in
\Lambda_L} \sigma_{x-a}\sigma_{x-a}'=Q_{\cal J}.
\end{equation}
Thus the overlap distribution function $P_{{\cal J}^a}=P_{\cal J}$ for
almost every ${\cal J}$ and all $a\in Z^d$; i.e., $P_{\cal J}$ 
{\it is translation-invariant\/}. 

As in the case of the translation-covariance of $\rho_{\cal J}$,
the translation-invariance of $P_{\cal J}$ has the conceptual
significance that a natural object like the Parisi order parameter
distribution should not (and in our construction does not) depend on
the choice of an origin.  But it also has the very important technical
(and physical) significance that $P_{\cal J}$ {\it must be self-averaged\/} because,
as already noted in Section~\ref{sec:EA} (in the discussion on
the number of pure states), a (measurable) translation-invariant function of 
random variables whose distribution is translation-ergodic is a constant
almost surely, by the ergodic theorem. (We remark that the fact that
we are dealing with a function of ${\cal J}$ whose value for each ${\cal J}$ is
an entire distribution is not a problem, since any
particular moment of that distribution is a real-valued function of ${\cal J}$.)

This answers the second part of our question:  the overlap distribution function
$P_{\cal J}(q)=P(q)$ is independent of ${\cal J}$.  It therefore
does not exhibit non-self-averaging (property (1)), and can exhibit {\it at
most\/} one of the two properties (2) and (3) discussed at the
end of Section~\ref{sec:SK}.  While property (2) (discreteness
of the locations of the delta-functions which appear in $P(q)$)
is not rigorously ruled out, it now seems like a highly implausible
possibility, since it would imply that the locations (and weights)
of the delta-functions (and consequently the values of $q$ which
correspond to {\it no\/} overlap value) are all independent of ${\cal J}$.
If property (2) is then eliminated as a realistic
possibility, then one can also rule out property (4) of the
SK picture, i.e., ultrametricity of all of the pure state overlaps for
fixed ${\cal J}$ \cite{NS96b}.  

Consequently, we have proved that {\it the standard
SK picture cannot be valid in any dimension and at any temperature\/}.
This result goes beyond our specific construction of the
Gibbs state $\rho_{{\cal J}}$ and overlap distribution $P_{{\cal J}}$, 
since {\it any\/} infinite-volume
translation-invariant overlap distribution function would be
self-averaging.  It would be quite peculiar if the overlap distribution
depended on the choice of origin of the coordinate system, and
we therefore regard the property of translation-invariance for
$P_{{\cal J}}$ (or translation-covariance for $\rho_{\cal J}$) as not specific to our
particular construction. 
 
Our conclusion is that {\it nearest-neighbor (and
in general realistic) spin glasses exhibit non-mean-field behavior,
because for those systems one can construct a non-self-averaged
Gibbs state\/} $\rho_{\cal J}$ {\it whose overlap distribution\/} $P_{\cal J}$ {\it is
self-averaged.\/}  The standard SK picture therefore cannot describe
realistic spin glasses at {\it any\/} dimension or temperature.
It is important to note that these conclusions apply to the {\it thermodynamics\/}
of spin glasses.  What might or might not occur in {\it finite\/} volumes involves
several subtle issues and will be discussed in
Section~{\ref{sec:nonstandard}.

While the demise of the standard SK~picture is interesting in itself,
and has important consequences for our understanding of spin glasses,
the methods used above and in our construction of $\rho_{{\cal J}}$ 
lead to perhaps more significant conclusions that might affect
our thinking on not only spin glasses, but disordered --- and
more widely, inhomogeneous --- systems at a deeper level.
Indeed, these methods indicate a path to a new and general 
approach for studying the thermodynamics of systems with many 
pure states.  One consequence will be the emergence
of a replacement for the standard SK picture, a new
picture which retains some mean-field flavor.  The general formulation
introduces several new concepts, among them replica non-independence
and a new definition of non-self-averaging, and relates them
to replica symmetry breaking, overlaps, and chaotic size dependence.
The unifying theme is the concept of the metastate, which is introduced
in the next section. Before that, we complete this section with 
a discussion of how we construct our thermodynamic state $\rho_{{\cal J}}$.

We begin by noting that we
cannot simply fix ${\cal J}$ and take an ordinary limit (i.e., through
a sequence of cube sizes $L$ chosen independently of ${\cal J}$)
of the finite cube, periodic
b.c. state $\rho_{{\cal J},\beta}^{(L)}$, as $L\to \infty$.  Unlike, say,
the $d=2$ homogeneous Ising ferromagnet, where such a limit exists (and equals
${1\over 2} \rho^+ +{1\over 2} \rho^-$) by spin flip symmetry considerations
(and the fact that $\rho^+$ and $\rho^-$ are the only pure states
\cite{A1980,H1981}), there is no guarantee for a spin glass that
there is a well defined limit.  (In fact, if such a limit
{\it does\/} exist for the spin glass, one can then prove \cite{NS92} 
that the {\it same\/} limiting 
state is obtained through the use of {\it antiperiodic\/} boundary
conditions --- a feature that already seems incompatible with an SK picture.)  

It is true that one can easily
prove, using compactness arguments, convergence along {\it subsequences\/} of
$L$'s for each ${\cal J}$.  But these subsequences should (in
an SK picture) be ${\cal
J}$-dependent.  The inconsistency between the existence of many
pure states and the existence of a 
thermodynamic limit for a sequence of finite-volume Gibbs states using
coupling-independent boundary conditions (such as periodic) 
and cube sizes has serious consequences not only for spin glasses
but also for systems in general with many competing states.
It suggests in the present case that, if many pure states exist, 
such a sequence of finite-volume
Gibbs state exhibits {\it chaotic size dependence\/} (CSD) and does
not converge to a limit.  The convergent (${\cal J}$-dependent) subsequences
would then give rise to different (non-self-averaged) pure states 
or mixed states with no way to make a (measurable) choice of a limit
state for each ${\cal J}$.

There is however a natural limit procedure which does give rise to an
infinite-volume Gibbs state $\rho_{\cal J}$, while avoiding such difficulties. Here, 
one considers the {\it joint\/}
distribution on the spins {\it and\/} the couplings; i.e., one
considers the distribution $\nu({\cal J}) \times \rho_{{\cal J},\beta}^{(L)}$
on the periodic cube $\Lambda_L$ \cite{NS96b}.  Then (again using
compactness arguments) some subsequence of $L$'s converges to
a limiting infinite-volume joint distribution 
$\mu({\cal J},\sigma)$. From this joint
distribution, $\rho_{\cal J}$ results when the spin configurations 
are chosen conditioned
on ${\cal J}$, which is chosen from $\nu({\cal J})$ in the usual way;
i.e., $\rho_{\cal J}$ is determined by the identity
$\mu({\cal J},\sigma) = \nu({\cal J})\times\rho_{\cal J}(\sigma)$.
The important difference with the earlier limit 
procedure is that this
one is valid for almost every ${\cal J}$, i.e., the subsequence of
$L$'s is ${\cal J}$-{\it independent\/}.  (The discussion so far has been
based on mathematically rigorous arguments. At this point however, we
would suggest --- but cannot rigorously prove --- that it is 
probably the case that 
using a subsequence of $L$'s is not needed for convergence, because the
use of a joint distribution for ${\cal J}$ and $\sigma$ should avoid CSD.)
A proof that the resulting limiting distribution
is indeed a Gibbs state may be found in \cite{AW,NSBerlin,Zurich}.
We note that such joint distribution limits were
considered, implicitly or explicitly, in Refs.~\cite{GKN,Le1977,Co1989,S1995}.

To get a clearer idea of this construction, consider the
following procedure.  Start with three cubes (labelled
a, b and c), all centered
at the origin, with volumes $L_a^d$, $L_b^d$, and $L_c^d$,
with $1\ll L_a \ll L_b \ll L_c$.  On the outermost box we
impose periodic boundary conditions.  The couplings are
fixed inside the intermediate box (and averaged over
between the intermediate and largest box); and in the
innermost box the overlap computation is done.  The
average over couplings between the intermediate and large
boxes is equivalent to an average over many boundary conditions
(consistent with the outer periodic b.c.) on the boundary of the
intermediate box.

Now let $L_c\to\infty$ while keeping $L_a$ and $L_b$ fixed;
then let $L_b\to\infty$, while still keeping $L_a$ fixed.
This sends our ``average over boundary conditions'' off to infinity
and results in an infinite-volume $\rho_{\cal J}$ which is
conditioned on {\it all\/} of the couplings and is
therefore non-self-averaged.  Finally, let $L_a\to\infty$;
this gives finally the overlap distribution $P_{\cal J}(q)$
between infinite-volume pure states appearing in the $\rho_{\cal J}$.

It is important to note that any analogue of this procedure 
for the SK model will result in an infinite-spin
Gibbs state, but a trivial one; i.e., it will already be 
self-averaged and therefore uninteresting.  This is because
fixing only finitely many couplings in the SK model and averaging
over the remainder is equivalent to averaging over {\it all\/}
of the couplings when $N\to\infty$.  This difference
between finite-dimensional and mean-field models is crucial.

We conclude by pointing out why our
construction (for the EA model) of 
the limiting joint distribution $\mu({\cal J},\sigma)$ 
yields translation-covariance for $\rho_{{\cal J}}$.
This is so because taking periodic
b.c.'s on the cube $\Lambda_L$ really means that the couplings and spins are 
defined on a (discrete) torus of size $L$, with a finite-volume joint distribution
invariant under torus translations.  This implies that any
(subsequence) limit joint distribution is
invariant under translations of $Z^d$, which in turn implies that
the infinite-volume Gibbs state $\rho_{\cal J}$ is translation-covariant. 
In the next section, we go beyond the construction of a single
limiting thermodynamic state by introducing the notion of metastates.

\section{Chaotic Size Dependence and Metastates}
\label{sec:meta}

In this section we will describe a new approach, introduced
in Ref.~{\cite{NS96c}, to studying
inhomogeneous and other systems with many competing pure states.
This approach is based on an analogy to chaotic dynamical
systems, and involves the replacement of the study of
a single thermodynamic state with an {\it ensemble\/}
of (pure or mixed) thermodynamic states.

In the previous section we were forced (by chaotic size
dependence) to replace a simple sequence of states on cubes with
periodic boundary conditions with a more complicated
sequence which involved an averaging over boundary
conditions, followed by sending this average off to
infinity.  This avoids chaotic size dependence
(at least for a ${\cal J}$-independent subsequence of volumes,
but probably altogether).  In this section, we will
pursue the opposite strategy --- we will forego the
end run around CSD, and instead use it to gather maximal
information about the disordered system.  The price
will be to abandon the usual procedure of constructing and studying
a single infinite volume Gibbs state $\rho_{\cal J}$.

The central observation behind this is that, at any
(large) fixed $L$ (and with periodic boundary conditions), 
the existence of multiple pure states should
generally require an approximate decomposition as
in Eq.~(\ref{eq:sum}) (see also (\ref{eq:sumea})):
\begin{equation}
\label{eq:sumL}
\rho_{{\cal J}}^{(L)}(\sigma)\approx\sum_\alpha 
W_{{\cal J},L}^\alpha\rho_{\cal J}^\alpha (\sigma)\ .
\end{equation}
For each $L$, the pure states appearing with the largest weights in the
sum will be those whose configurations 
within the volume of size $L$ are best adapted to the boundary
condition.  Chaotic size dependence requires that the pure
states and weights appearing within 
this decomposition vary persistently as $L$
is increased (though it says nothing about the rate at which this
variation occurs).

The analogy with the chaotic orbit of a dynamical system 
follows from the identification of cube size $L$ with
time $t$ along such an orbit.  A (chaotic) dynamical system's trajectory 
will appear random if one considers the sequence of points 
along its orbit, but one can describe its long-time
behavior by studying the appropriate probability measure $\kappa_{dyn}$
on state space.  That is, one can construct a histogram
at times $t_1, t_2, \ldots, t_N$ with $N$ increasing to 
infinity, and study the fraction of times spent by the orbit
in different parts of state space (in a continuous space
this would require breaking the space up into bins).  The
$N\to\infty$ limit of this process yields a well-defined
$\kappa_{dyn}$.

Similarly, we consider at fixed ${\cal J}$ a histogram
of finite volume Gibbs states $\rho_{{\cal J}}^{(L_1)},\rho_{{\cal J}}^{(L_2)},\ldots,
\rho_{{\cal J}}^{(L_N)}\to\kappa_{\cal J}$ as $N\to\infty$.  The information
contained in $\kappa_{\cal J}$ provides the fraction of cube sizes
which the system spends in different (possibly mixed) thermodynamic states
$\Gamma$.  We refer to $\kappa_{\cal J}$, which is a probability measure
on thermodynamic states $\Gamma$ at fixed ${\cal J}$, as the {\it metastate\/}.

To simplify notation, it will be assumed in the ensuing discussion
that convergence to the metastate is valid without need for a subsequence
of $N$'s or a subsequence $L_1, L_2, \ldots$ of the cube sizes.
We point out however that K\"ulske has studied some models, e.g. the
Curie-Weiss random-field model, in which choosing a sparse subsequence
of sizes is necessary for the empirical distribution
(i.e., the histogram)
to converge (for almost every disorder configuration) 
to the metastate.  We will not discuss these
issues here, but refer the reader to Ref.~\cite{Ku96} for details.

Our empirical distribution approach to construction of a metastate, 
based on CSD for fixed ${\cal J}$, constructs $\kappa_{\cal J}$ as 
the limit of $\kappa_{\cal J}^{(L)}$, 
a type of microcanonical ensemble in which each of the finite volume states
$\rho_{\cal J}^{(1)},\rho_{\cal J}^{(2)},\ldots,\rho_{\cal J}^{(L)}$
has weight $L^{-1}$.  This limit can be understood in the following
way:  consider a (nice) function on the states, such as the correlation
$\langle\sigma_x\sigma_y\rangle(\cdot)$; i.e., 
$\langle\sigma_x\sigma_y\rangle(\Gamma)$ is this correlation computed
using a particular infinite-volume Gibbs distribution $\Gamma$,
and $\langle\sigma_x\sigma_y\rangle(\rho^{(L)})$ is the same
correlation computed using the finite volume Gibbs distribution
$\rho^{(L)}$ (we suppress the ${\cal J}$ index here, which is understood).
If $\left[\cdot\right]_\kappa$ denotes an average of a state-dependent
function over the {\it metastate\/} (i.e., the function of each state
is weighted using the weight of the corresponding state within the
metastate), then
\begin{equation}
\label{eq:corrs}
\lim_{L\to\infty} (1/L)\sum_{l=1}^L\langle\sigma_x\sigma_y\rangle(\rho^{(l)})
=\left[\langle\sigma_x\sigma_y\rangle(\Gamma)\right]_\kappa\quad .
\end{equation}
Furthermore such an equation similarly holds for any other (nice) function
of finitely many correlations (regarded as a function on states).

There is another approach to constructing the metastate, due to Aizenman
and Wehr \cite{AW}, which uses the randomness of the couplings
directly, in a manner similar to that of the construction 
of $\rho_{\cal J}$ in
the previous section.  There we studied
the limiting joint distribution $\mu({\cal J},\sigma)$ of the random
pairs $({\cal J},\sigma^{(L)})$ distributed for finite $L$ by 
$\nu({\cal J})\times\rho_{{\cal J}}^{(L)}$.  Here one considers instead the
random pair $({\cal J},\rho_{{\cal J}}^{(L)})$ at finite $L$.  We will
not discuss various technicalities associated with this
method; details can be found in Refs.~\cite{AW,NSBerlin,Zurich}.  We will simply note here that
the two approaches (at the very least along common 
${\cal J}$-independent subsequences) yield the same limiting metastate.

The metastate $\kappa_{\cal J}$ contains all of the thermodynamic information
about a system, in this case the EA spin glass with
coupling realization ${\cal J}$.  As such, it contains far more information
than the single thermodynamic state $\rho_{\cal J}$ generated by the
construction of the previous section (or any other single state).
In fact, it can be seen \cite{AW,NSBerlin,Zurich} that the 
$\rho_{\cal J}$ of the previous section is the {\it average\/} thermodynamic state 
of the ensemble of states within the metastate $\kappa_{\cal J}$,
in the following sense:  consider any spin correlation
in the state $\rho_{\cal J}$, e.g., $\langle\sigma_{x_1}
\cdots\sigma_{x_n}\rangle_{\rho_{\cal J}}$.  This equals the average
(over the metastate) of the correlation function of the same set of spins over
{\it all\/} thermodynamic states $\Gamma$ of the ensemble.  So if
$\kappa_{\cal J}(\Gamma)\ d\Gamma$ denotes (formally) the probability of appearance
of the states within a region of state space centered on
$\Gamma$ of state-space volume $d\Gamma$, then 
\begin{equation}
\label{eq:barycenter}
\langle\sigma_{x_1}\cdots\sigma_{x_n}\rangle_{\rho_{\cal J}}
=[\langle\sigma_{x_1}\cdots\sigma_{x_n}\rangle_\Gamma]_{\kappa_{\cal J}}
=\int \langle\sigma_{x_1}\cdots\sigma_{x_n}\rangle_\Gamma\ 
\kappa_{\cal J}(\Gamma)\ d\Gamma
\end{equation}
and similarly for all other correlation functions.

We see that one problem with the standard SK picture (and with
other standard thermodynamic treatments of systems
with many competing states) is that the state
$\rho_{\cal J}$ (or any other single state, pure or mixed)
is simply not a rich enough description of the
$L\to\infty$ behavior of a thermodynamic
system where CSD occurs.  In these approaches, one
is in effect replacing with a single average
all of the information contained
in an entire distribution.

To illustrate the nature of the metastate, we 
now present some simple examples.

The first is the trivial case of a single
pure phase, e.g., the paramagnetic state.
Then $\lim_{L\to\infty}\rho_{{\cal J}}^{(L)} = \rho_{\cal J}$
is a single pure state, there is no CSD,
and $\kappa_{\cal J}(\Gamma)=\delta(\Gamma-\rho_{\cal J})$.

In the second example we suppose that the scaling/droplet
picture is correct, so that only two pure
states $\rho'_{\cal J}$ and $\rho''_{\cal J}$ exist,
related by a global spin flip.  Then (as in
the $d=2$ homogeneous Ising ferromagnet with periodic
or free b.c.'s)
\begin{equation}
\label{eq:twostates}
\lim_{L\to\infty}\rho_{{\cal J}}^{(L)} =  {1\over 2}\rho'_{\cal J}+{1\over 2}\rho''_{\cal J}
\end{equation}
and again there would be no CSD.  Indeed, the analogy here
is to a dynamical system with a simple fixed point.
The metastate is simply
\begin{equation}
\label{eq:metatwostates}
\kappa_{\cal J}(\Gamma)=\delta\left(\Gamma-[{1\over 2}\rho'_{\cal J}+
{1\over 2}\rho''_{\cal J}]\right)\quad .
\end{equation}

However, we can introduce a slight variation of this
procedure to illustrate the potential sensitivity
of the metastate to the boundary conditions used in
the limiting procedure.  Suppose that, instead of
periodic boundary conditions for each $L$, 
we use {\it fixed\/} boundary conditions,
e.g., all spins are $+1$ on the boundary of each
$L^d$ cube appearing in the sequence.  This of course
breaks the spin flip symmetry, and for some $L$'s
the state $\rho'_{\cal J}$ will be preferred, while others will
prefer $\rho''_{\cal J}$, depending in each case on whether the 
sum along the boundary of $\langle \sigma_x \rangle_{\rho_{{\cal J}}^\prime}$
is (substantially) positive or negative.
(There may be occasional $L$'s where
the preference for each state is roughly
equal, but this should be a negligible fraction of $L$'s and
so wouldn't show up in the limiting histogram which yields the
metastate.)

So in this case we get chaotic size dependence, albeit 
of a trivial sort:  $\rho_{{\cal J}}^{(L)}\approx\rho'_{\cal J}$
for roughly half of the $L$'s, and $\rho_{{\cal J}}^{(L)}\approx\rho''_{\cal J}$
for the remainder.
The metastate is now
\begin{equation}
\label{eq:metatwostatesdisp}
\kappa_{\cal J}(\Gamma)={1\over 2}\delta(\Gamma-\rho'_{\cal J})+{1\over
2}\delta(\Gamma-\rho''_{\cal J})\quad .
\end{equation}
Here the metastate is a rough analogue to the $\kappa_{dyn}$ obtained
from a discrete time dynamical system with an attractive orbit of
period two.

This is our first example in which the metastate is not
simply a delta-function on thermodynamic states.  We
call this behavior {\it dispersal of the metastate\/},
and it is intimately connected with CSD.  From this
and the previous example, it should be clear that dispersal
of the metastate is quite different from the mere
existence of multiple states; while the existence of more
than one state is necessary for dispersal to occur,
it by no means guarantees it.  

The above discussion leads naturally to the following
possibility, first proposed in \cite{NS96c} as a
possible candidate for the EA metastate, based in
turn upon earlier work in \cite{NS94}.  Suppose
that the EA spin glass has many pure states
in some $d$ and at some $\beta$, but unlike
in the mean-field picture each volume ``sees'' essentially only
one pair at a time.  In other words, for every
$L$ (and once again using periodic boundary conditions), one finds that
\begin{equation}
\label{eq:possfive}
\rho_{{\cal J}}^{(L)}\approx {1\over 2}\rho_{\cal J}^{\alpha_L}+{1\over 2}\rho_{\cal J}^{-\alpha_L}
\end{equation}
where $-\alpha$ refers to the global spin flip of pure state $\alpha$.
In any volume, this looks very much like the droplet/scaling picture,
but its thermodynamic behavior is considerably different:  there
are infinitely many pure states and which pair appears in any finite
volume depends chaotically on $L$.  Unlike the droplet/scaling
picture, this new possibility exhibits CSD and dispersal of the metastate. 
In this ``chaotic pairs'' picture the (periodic b.c.) metastate is dispersed over (infinitely) many
$\Gamma$'s, of the form $\Gamma =
\Gamma^\alpha={1\over 2}\rho_{\cal J}^\alpha+{1\over 2}\rho_{\cal J}^{-\alpha}$.

It is interesting to note that just this type of behavior
is observed for the many ground states of a simple highly
disordered spin glass model in high dimension(\cite{NS94}; see also \cite{CMB94}).
(Indeed, for the EA model itself, we would expect this
type of behavior to occur at $T=0$ if infinitely many
ground states exist.)

One can study metastates in models more complicated than those
above, but still simpler than the EA spin glass,
and a discussion of some of these (e.g., random-field Ising
models, the highly disordered spin glass of Ref.~\cite{NS94},
and the homogeneous XY model with random b.c.'s) appears in \cite{NSBerlin},
to which the interested reader is referred for details.
At this point, however, we will proceed and use the
ideas introduced in this chapter to revisit the concepts
of replica symmetry breaking and non-self-averaging,
and will introduce some new concepts such as replica non-independence.
The idea of the metastate will enable us to relate, explain,
and unify these concepts.  We will then return to the EA model
and discuss the remaining possibilities for its metastate, and
therefore its low temperature thermodynamic structure.

\section{Replica Symmetry Breaking, Replica Non-Independence, and Overlap Distributions}
\label{sec:rsbetc}

We have discussed in Section~\ref{sec:standard} the Parisi order
parameter distribution $P_{\cal J}(q)$ in the standard SK picture,
whose counterpart $P_{{\cal J},N}(q)$ for the SK model was successful in describing
mean-field spin glass ordering.  In the standard SK model
$P_{\cal J}(q)$ is constructed as the distribution of
the overlap random variable $Q$, which in turn
is constructed according to Eq.~(\ref{eq:overlapea}), where
the spin configurations $\sigma$ and $\sigma'$ are
chosen from the product distribution 
$\rho_{\cal J}(\sigma)\rho_{\cal J}(\sigma')$; i.e., 
each is chosen independently from the same
(thermodynamic) state $\rho_{\cal J}$.  

But now, given the metastate point of view discussed
in the previous section, we know that the state
$\rho_{\cal J}(\sigma)$ is really the average over
the metastate, in the sense described by Eq.~(\ref{eq:barycenter}).
Equivalently,
\begin{equation}
\label{eq:rhoav}
\rho_{{\cal J}}(\sigma)=\int\ \Gamma(\sigma)\kappa_{\cal J}(\Gamma)\ d\Gamma\quad .
\end{equation}

So each time a pair of spin configurations, say $(\sigma^1 , \sigma^2)$,
is chosen from $\rho_{{\cal J}}(\sigma^1)\rho_{{\cal J}}(\sigma^2)$,
an independent $\Gamma$ is used for each configuration. I.e.,
$\sigma^1$ is chosen from $\Gamma^1$ and
$\sigma^2$ from $\Gamma^2$ with $(\Gamma^1,\Gamma^2)$ chosen
from $\kappa_{{\cal J}}(\Gamma^1)\kappa_{{\cal J}}(\Gamma^2)$;
$\Gamma^1$ and $\Gamma^2$ will in general be distinct if
the metastate is dispersed.
This in turn means in essence (see (\ref{eq:corrs}))
that the spin configuration
$\sigma^1$ is chosen using the
distribution $\rho_{\cal J}^{(L_1)}$ and 
$\sigma^2$ from $\rho_{\cal J}^{(L_2)}$,
with $L_1\ne L_2$.  It seems more natural instead
to take the two replicas from the same
distribution, i.e., for a {\it single\/} $L$,
and therefore from the {\it same\/} $\Gamma$.  
As Guerra has pointed out \cite{Guerra},
this order of operations (in which replicas are 
taken {\it before\/} $L \to \infty$) could yield
a different result than that obtained
by first letting $L\to\infty$ to obtain
an infinite volume state $\rho_{\cal J}$ and
then taking replicas.  The noncommutativity of these operations
will be shown to follow from a phenomenon we call
{\it replica non-independence\/},
which is not the same as replica symmetry
breaking, as will be seen below.
But first we will explore the meaning
of this new way of taking replicas.

Taking replicas first (i.e., from the same $L$) really means,
in terms of the metastate, that they 
are being taken from the same $\Gamma$,
i.e., from $\Gamma(\sigma^1)\Gamma(\sigma^2)$
for some $\Gamma$ chosen from
$\kappa_{\cal J}(\Gamma)$.  (Without metastates,
it would be difficult to assign a clear
meaning to this statement.)  For $n$ replicas
(where $n$ can be any positive integer, or infinity)
we take $n$ uncoupled (but identical)
Hamiltonians (and boundary conditions) 
in the cube $L^d$.
We then use for finite $L$ the product measure
$\rho_{\cal J}^{n(L)} =
\rho_{\cal J}^{(L)}(\sigma^{1(L)})\rho_{\cal J}^{(L)}(\sigma^{2(L)})
\ldots\rho_{\cal J}^{(L)}(\sigma^{n(L)})$.  The limiting
joint distribution for $({\cal J},\sigma^{1(L)},
\sigma^{2(L)},\ldots\sigma^{n(L)})$ is then
of the form $\nu({{\cal J}})\rho_{\cal J}^n(\sigma^1,\sigma^2,\ldots\sigma^n)$
for some $\rho_{\cal J}^n$ that we call the infinite volume $n$-replica measure.
(The mathematical analysis of this limit procedure is
essentially the same as was discussed for $n=1$ in Section IV above
and, with more detail, in \cite{NSBerlin,Zurich}.)

In this approach, replicas in the infinite volume limit are described
by $\Gamma(\sigma^1)\Gamma(\sigma^2)\ldots$ with $\Gamma$
distributed by $\kappa_{\cal J}$; replicas in finite
volume are taken from the {\it same\/} $L$, and
$\kappa_{\cal J}$ describes the sampling of states as $L$ varies.

A crucial point, as emphasized by Guerra \cite{Guerra},
is that {\it in the infinite volume replica measure\/} $\rho_{\cal J}^n$,
{\it the replicas need not be independent\/}, although they
of course {\it are\/} independent in the finite volume measure
$\rho_{\cal J}^{n(L)}$.  The replicas in infinite volume can be thought of as 
coupled through ``boundary conditions at infinity''.

If this occurs, we say \cite{NS96c} that the system
displays replica non-independence (RNI).  The
presence of RNI means that $\rho_{\cal J}^n$, which is a thermodynamic state
for the uncoupled replica Hamiltonians, is {\it not\/}
simply equal to the product of the individual Gibbs states 
$\rho_{\cal J}(\sigma^i)$.  
In general, we have from the above description that
\begin{equation}
\label{eq:pjn}
\rho_{\cal J}^n(\sigma^1,\sigma^2,\ldots\sigma^n)
=\int\left[\Gamma(\sigma^1)\Gamma(\sigma^2)\ldots\Gamma(\sigma^n)\right]
\kappa_{\cal J}(\Gamma)\ d\Gamma
\end{equation}
This makes it clear that RNI is equivalent to dispersal of the
metastate.  If the metastate is nondispersed, its weight
is concentrated entirely on a single thermodynamic state,
so $\kappa_{\cal J}$ is a delta function, and the RHS of
Eq.~(\ref{eq:pjn}) reduces to a simple product of 
Gibbs states (each of which is the
single state on which the metastate is 
concentrated).  Otherwise, the product
decomposition of $\rho_{\cal J}^n$ is as a mixture
over $\kappa_{\cal J}$.  This also shows that
RNI is equivalent to the non-interchangeability
of the operations of taking replicas and the
thermodynamic limit.

In Ref.~\cite{NS96c}, these points were explained
using the idea of ``metacorrelations''.  Just as
the usual correlations
$\langle\sigma_{i_1}\cdots\sigma_{i_m}\rangle_\Gamma$
are moments (in this case, of order $m$) characterizing
the thermodynamic state $\Gamma$, metacorrelations
are moments that characterize the metastate $\kappa$.
I.e., they are the averages (over the metastate) of functions
$g(\Gamma)$ on the states that are monomials (of order $m$)
in the correlations (of varying orders):
\begin{equation}
\label{eq:corr}
[g(\Gamma)]_{\kappa} = \left[\langle\sigma_{A_1}\rangle_{\Gamma}\dots
\langle\sigma_{A_m}\rangle_{\Gamma}\right]_{\kappa}\quad ,
\end{equation}
where $\sigma_A$ denotes
$\sigma_{i_1}\dots\sigma_{i_k}$ for the set $A = \{i_1,\dots,i_k\}$.
As noted in Ref.~\cite{NS96c}, restriction to
metacorrelations of order $m=1$ yields $\rho_{\cal J}$, to $m\le 2$
yields the two-replica measure $\rho_{\cal J}^2(\sigma^1,\sigma^2)$,
which corresponds to ``integrating out'' all the other replicas in
$\rho_{\cal J}^\infty$, and so on.  The measure $\rho_{\cal J}^\infty$ therefore
not only contains information about arbitrarily many replicas, but
since it determines all the metacorrelations, it
also contains all information about the metastate $\kappa_{\cal J}$.

Replica symmetry breaking (RSB) occurs when individual
thermodynamic states $\Gamma$ 
(chosen from $\kappa_{\cal J}$) are mixtures of multiple
pure states, so that even, when restricted to a single
$\Gamma$, replicas can come from different pure
states, in the spirit of the Parisi {\it ansatz\/}.
This definition allows for what we call trivial RSB (e.g., in a two-state
picture), but corresponds to the more familiar
meaning when many pure states are present in $\Gamma$. 
The presence of RSB means that 
when one decomposes each mixed $\Gamma$ in 
Eq.~(\ref{eq:pjn}) into pure states,
then the permutation symmetry between 
different replicas is lost in each of the products
where a pure state is chosen for each replica. 
It follows that RSB and RNI are distinct phenomena,
and either can occur without the other.

Although we have a new way of constructing a replica
measure, we can still take overlaps in the usual way,
i.e., according to Eq.~(\ref{eq:overlapea}).
The distribution of an overlap $Q$, though, depends on
how $\sigma$ and $\sigma'$ are chosen.  Because
of the possibility of RNI, we no longer take overlaps
(between one or more pairs of replicas)
from the product measure 
$\rho_{\cal J}(\sigma^1)\rho_{\cal J}(\sigma^2)\dots$, but instead from
the more suitable replica measure $\rho_{\cal J}^\infty$.
Because of this, the nature of the overlaps changes.
For example, the distribution of a single overlap $Q$
is no longer the $P_{\cal J}(q)$ obtained from
$\rho_{\cal J}(\sigma)\rho_{\cal J}(\sigma')$
but rather is $\int P_{\Gamma}(q)\kappa_{\cal J}(\Gamma)d\Gamma$,
where $P_{\Gamma}(q)$ denotes the overlap distribution
obtained from $\Gamma(\sigma)\Gamma(\sigma')$.
When $\kappa_{\cal J}$ is dispersed, $P_\Gamma$ may or may not
depend on the $\Gamma$ chosen from $\kappa_{\cal J}$.
(It does not in the chaotic pairs picture but does in the 
nonstandard SK picture discussed below.) Information
on this dependence is contained in the overall ``overlap
structure'', by which we mean the joint distribution
of all overlaps $Q^{ij}$ betwen all pairs of replicas
$(\sigma^i,\sigma^j)$ from $\rho_{\cal J}^\infty$. This (possible)
dependence on $\Gamma$ is significant because, as in our
analysis above of the standard SK picture, 
{\it the overlap
structure still does not depend on\/} ${\cal J}$,
by essentially the same arguments using
translation-invariance of the overlaps
and translation-ergodicity of the
coupling distribution $\nu({\cal J})$.  
More specifically, regarding $P_\Gamma$ as random because
of its dependence on $\Gamma$ for fixed ${\cal J}$, 
the probability of appearance of a particular set of weights {\it and\/}
corresponding locations of the overlap delta-functions will not depend on ${\cal J}$. 
(Here, we are describing the situation, discussed at length
in the next section of the paper, in which $P_\Gamma$,
for each $\Gamma$, has an SK type form.)

In realistic systems, thermodynamic state observables can depend on the
bulk couplings and/or on the couplings at infinity. Thus
we observe that {\it there are two
distinct types of dependence:
(i) on ${\cal J}$, and (ii) on the state $\Gamma$ within the 
metastate $\kappa$ for fixed ${\cal J}$\/}. 
We have seen that replica overlaps cannot have the first type of dependence,
but can in principle have the second kind.  In that case, if one examines
the same (finite) volume for two different coupling realizations,
it could happen that two different sets
of weights and overlap locations are seen (in the approximate sense
corresponding to the fact that we're restricted here to finite volumes,
so that, e.g., the finite volume overlap distribution
$P_{\cal J}^{(L)}$ is not a sum of exact delta functions).  
It is logically possible that 
in such a case, fluctuations in $P_{\cal J}^{(L)}$ persist
for arbitrarily large $L$.  From our previous discussions, however, it
would be {\it incorrect\/} to conclude that there is an infinite volume 
overlap distribution that is non-self-averaging (i.e.,
that depends on ${\cal J}$). Rather, it would imply
that the limit $\lim_{L\to\infty} P_{\cal J}^{(L)}$ does not exist;
i.e., that $P_{\cal J}^{(L)}$ exhibits chaotic size dependence \cite{Noteoverlaps}.

Our conclusion is that if overlap fluctuations
(due to coupling dependence) don't vanish 
as $L\to\infty$ \cite{PS}, this does {\it not\/} mean
that the standard SK picture of overlap non-self-averaging 
holds; rather, it is a signal that the {\it second\/} kind
of dependence holds for infinite volume.  

With the new approach outlined above, a replacement
for the standard SK picture suggests itself.
This replacement at first may seem very unusual
and different from previous understandings of
thermodynamic spin glass structure, but it
falls out naturally from the ideas presented in this
and the previous section.  In the following section
we ask, how can at least some of the familiar
characteristics of the Parisi version of spin glass
ordering be retained in realistic spin glasses?
We will see that the ``maximal'' mean-field
picture allowed, given our new understanding of
metastates and their consequences, has an intricate
and novel thermodynamic structure.

\section{The Nonstandard SK Picture}
\label{sec:nonstandard}

We saw in Section~\ref{sec:standard} that the familiar
thermodynamic picture usually associated with the Parisi {\it ansatz\/}
applied to the EA model, which we called the
standard SK picture, could not be valid in any
dimension and at any temperature.  Any thermodynamic
theory of realistic spin glasses will differ considerably
from this picture.  The question then is whether and
how any aspects of mean-field behavior can survive
in such a theory.  We now address this question.

We begin by asking what a maximal mean-field
picture would look like in {\it finite\/} volume.
There have been a number of numerical simulations 
(e.g., \cite{Caracciolo,MPRR,RBY}) which appear
to see a Parisi-like structure of finite-volume
states, i.e., the appearance of several states
with nonnegligible weight, several (approximate) delta-function
overlaps whose positions depend on coupling realization, 
and a Parisi-like $P(q)$  (i.e., delta-functions
at $\pm q_{EA}$ connected by a continuous part) after 
averaging over the couplings.  (See, however, \cite{FHcrit}
for a criticism of \cite{Caracciolo}.)  We will not attempt
to resolve controversies associated with these or other
simulations, nor will we speculate whether, if correct,
these results persist for larger volumes.  Rather,
we ask {\it if\/} such results should hold
for arbitrarily large volumes, what does
that imply about the thermodynamics of spin glasses,
given that the usual thermodynamic extrapolation
of these finite-volume results (i.e., the standard
SK picture) is incorrect?

We will see that the metastate approach allows
us to construct such a thermodynamic scenario,
which we call the {\it nonstandard SK picture\/}.
This picture, or one closely related to it,
must describe the thermodynamics of realistic spin
glasses if the above {\it finite\/}-volume picture
is correct.  That is, the nonstandard
SK picture allows for properties (1) -- (4)
appearing at the end of Section~\ref{sec:SK}
(or more precisely, finite volume versions
of these properties)
to hold in any fixed finite volume (with, e.g.,
periodic boundary conditions).  It is
therefore a maximal mean-field picture,
as promised at the end of the previous
section.  However, the thermodynamics
to which it gives rise is
novel and unconventional.  It displays
RSB and RNI (equivalently, CSD) and
a type of non-self-averaging, suitably
redefined (as described in the previous section).
It does not have the features commonly thought
of as associated with the Parisi {\it ansatz\/},
e.g., ultrametricity of all of the pure states
\cite{VHO,BCPR}, but displays some of its properties
in a more limited fashion.

As a starting point, then, we require that in
any (large) finite volume, the Gibbs state is
an approximate decomposition over many
pure states:
\begin{equation}
\label{eq:sumfinite}
\rho_{{\cal J}}^{(L)}(\sigma)\approx\sum_
\alpha W_{\cal J}^{\alpha,L}\rho_{\cal J}^\alpha (\sigma)\ ,
\end{equation}
where a few states dominate the sum.  From the metastate
point of view, this implies that each $\Gamma$ 
(chosen from $\kappa_{\cal J}$) is a
mixed state with a nontrivial decomposition into
pure states, namely
\begin{equation}
\label{eq:sumgamma}
\Gamma=\sum_\alpha W_{\Gamma}^{\alpha}\rho_{\cal J}^\alpha (\sigma)\ ,
\end{equation}
and this decomposition is discrete but with many nonzero
weights $W_\Gamma^\alpha$ \cite{Notediscrete}.

In order that this scenario correspond to the usual
expectations of the Parisi/SK picture in {\it finite\/}
volumes (and at fixed ${\cal J}$), we require that the fixed-$\Gamma$ overlap
distribution
\begin{equation}
\label{eq:PJgamma}
P_{\Gamma}(q)=\sum_{\alpha,\gamma}W_{\Gamma}^\alpha W_{\Gamma}^\gamma
\delta(q-q^{\alpha\gamma})\quad 
\end{equation}
display the form consistent with property (2) listed
at the end of Section~\ref{sec:SK}, that is, a countable sum of
many delta-functions.  (Note that the occurrence of many
(distinct) $q_{\alpha\gamma}$'s
is an additional requirement, and does not
follow automatically from Eq.~(\ref{eq:sumgamma}).)
The metastate must be an ensemble of many such $\Gamma$'s
(in fact, a continuum of them, as we explain below),
each of which yields a pair of delta-functions at $\pm q_{EA}$,
but with the locations of the remaining delta-functions
being $\Gamma$-dependent.  We further require that the
locations of the delta-functions within a specific
$P_\Gamma(q)$ be ultrametric.  

The above requirements are consistent with properties
(1) -- (4) of Section~\ref{sec:SK} holding for any finite
volume, including (conventional) non-self-averaging for arbitrarily
large $L$.  However, instead of the straightforward
extrapolation to infinite volumes characteristic
of the standard SK picture, the thermodynamic
properties of this nonstandard SK picture are
considerably different.  We now discuss what
these properties are.

The crucial conceptual point is that the translation-covariance
of the metastate $\kappa_{\cal J}$ still
requires that the resulting ensemble of overlap distributions
is independent of ${\cal J}$.  The metastate
in this picture must be an ensemble of many $\Gamma$'s
, with a single $\Gamma$ 
appearing in any fixed cube $L^d$ (with, e.g.,
periodic boundary conditions).  
The dependence on $\Gamma$ (as $\Gamma$ varies within
the metastate ensemble) is the new sort of 
non-self-averaging discussed at the end of
the previous section. It is
clear then that this picture must
have both nontrivial RSB (because each $\Gamma$
is a sum over many pure states), and CSD
(and RNI) since the metastate is dispersed. 

Finally, we require that the (averaged) Parisi order parameter
$P(q)$ have the usual form, that is, two
delta-functions at $\pm q_{EA}$, connected by a
continuous component with nonzero weight everywhere
in between; however, the averaging {\it must
now be done over the states\/} $\Gamma$ {\it within
the metastate\/} $\kappa_{\cal J}$, all at fixed
${\cal J}$, rather than over ${\cal J}$ itself:
\begin{equation}
\label{eq:Pgamma}
P(q)=[P_\Gamma(q)]_{\kappa_{\cal J}}=\int P_\Gamma(q)\kappa_{\cal J}(\Gamma)d\Gamma.
\end{equation}
In order for this requirement to be valid along with
discreteness of the individual $P_\Gamma$'s, it must 
be that there is a {\it continuum\/} of $\Gamma$'s
in the metastate ensemble. So we have replaced dependence
on coupling realization ${\cal J}$ with dependence on the state
$\Gamma$ within the metastate for {\it fixed\/} ${\cal J}$.

We see that the nonstandard
SK picture differs from the usual mean-field picture
in several important respects.  One is the lack of dependence
of overlap distributions on ${\cal J}$, and the replacement
of the usual sort of non-self-averaging with the concept of dependence
on states within the metastate.  Another important
difference is that, in the nonstandard SK
picture, a continuum of pure states {\it and\/}
their overlaps must be present; therefore,
{\it ultrametricity would not hold in general
among any three pure states chosen at fixed\/}
${\cal J}$, unlike in the standard SK picture
(see, for example, \cite{VHO,BCPR}).  (The
argument supporting this conclusion is
presented in Ref.~\cite{NS96b}.)  Rather,
the pure states at fixed ${\cal J}$ are split up
into (a continuum of) families,
where each family consists of those pure
states occuring in the decomposition of
a particular
$\Gamma$, and only within each such family 
would ultrametricity hold.  

We have presented the nonstandard SK picture as 
a replacement for the more standard mean-field
picture; if realistic spin glasses display
any mean-field features, something like it must
occur.  However, this leaves open the question
of what actually happens in realistic spin glasses.
In particular, does the nonstandard SK
picture actually occur?  It turns out to have an
important covariance property which may provide a clue.

For specificity, consider the EA model with a
(mean zero, variance one) Gaussian coupling distribution.
Suppose that we change a {\it finite\/} number of couplings.
The metastate $\kappa_{\cal J}(\Gamma)$,
in addition to translation-covariance, is also 
covariant with respect to this change \cite{AW}; that is,
the ensemble transforms (as would any probability measure)
under a change of variables, $\Gamma \to \Gamma'$.
Here, $\Gamma'$ is the thermodynamic state with correlations
$\langle \sigma_A \rangle_{\Gamma'} = 
\langle \sigma_A e^{-\beta \Delta H} 
\rangle_\Gamma / \langle e^{-\beta \Delta H} \rangle_\Gamma$,
where $\Delta H$ is the change in the Hamiltonian.
Under this change of variables,
pure states remain pure and their overlaps
don't change.  However, the weights which appear
in Eq.~(\ref{eq:PJgamma}) {\it will\/} in general change.
Nevertheless, the overall overlap structure
(i.e., the probability of appearance of 
a given set of weights and overlap locations)
must remain invariant.

We propose \cite{NSBerlin} this 
covariance property under coupling changes as an
appropriate analogue to that of of dynamical systems having a probability
measure invariant under the dynamics.  Our reasoning is as follows.
Suppose we consider free b.c.'s. Changing from a cube of
size $L$ to one of size $L+1$ corresponds to taking
a certain layer of couplings and changing them from
zero to nonzero values. Having already made an analogy
between $L$ and the time $t$ for the dynamical
system, it seems appropriate to extend it to one
between dynamical invariance ($t \to t+1$) and 
coupling covariance (${\cal J} \to {\cal J}+ \Delta {\cal J}$).
The analogy is even clearer if we consider increasing 
volumes not by a whole layer at a time but by
a single site at a time.  Exploitation of this
covariance property could result in a new
type of cavity method \cite{MPV,DT,MPV86} for studying
the properties of realistic spin glass models.

In the nonstandard SK picture, there
seems every reason to expect nontrivial dependence
of, e.g., $\langle e^{-\beta \Delta H} \rangle _{\cal J}^\alpha$
on the many $\alpha$'s appearing for each $\Gamma$. Thus,
under changes of finitely many couplings, each
$P_\Gamma$ would be changed to a $P_{\Gamma'}$ with the
same $q_{\alpha\gamma}$'s but with {\it different} weights.
Nevertheless, by the translation-invariance/ergodicity
argument mentioned earlier in this section, the
{\it distribution} of the $P_\Gamma$'s (as $\Gamma$
varies over the metastate) in fact does not depend
on ${\cal J}$ and hence is unchanged by
${\cal J} \to {\cal J}+ \Delta{\cal J}$.

Thus the above covariance property under changes
of couplings places a large number of constraints on the
distribution of the $P_\Gamma$'s that can arise in
the nonstandard SK picture. We wonder whether all 
these constraints (which do {\it not} arise 
either in the droplet/scaling or in the chaotic pairs pictures) can
actually be satisfied.  Clearly, more study of
this issue is needed.

\section{Conclusions}
\label{sec:conclusions}

We have shown that the traditional picture of spin
glass thermodynamics, based on the Parisi {\it ansatz\/}
as applied to finite-dimensional models, cannot hold for 
realistic spin glasses in any dimension and at any
temperature.  This standard SK picture is a natural
and straightforward extrapolation to infinite
volumes of the main features of spin glass ordering
uncovered by Parisi and others for the SK model.
It assumes a single infinite volume
overlap distribution function
$P_{\cal J}(q)$ which is non-self-averaging,
i.e., dependent on ${\cal J}$.  This picture proposes
that the pure states are chosen independently
from some mixed (and, of course, non-self-averaged) 
thermodynamic state $\rho_{\cal J}$ with a decomposition
of the form of Eq.~(\ref{eq:sumea}) and that the
resulting $P_{\cal J}(q)$ will
consist of (many) discrete delta-functions lying
between a pair at $\pm q_{EA}$.  The locations
of the delta-functions (except for
the pair at $\pm q_{EA}$) and their weights 
depend on the coupling
realization ${\cal J}$, but for any fixed ${\cal J}$ their locations
are ultrametric.  When averaged over the (uncountably many)
coupling realizations chosen from the coupling distribution,
the order parameter distribution $P(q)=\overline{P_{\cal J}(q)}$
shows the characteristic form of a continuous component
connecting the delta-functions at $\pm q_{EA}$, and nonzero
everywhere in between (at least for $0<T<T_c$). 

We have shown, however, that this picture can never
hold:  {\it any \/} $P_{\cal J}(q)$ {\it with the weak (and
physically reasonable) property of translation-invariance
must be self-averaging\/}, due to the underlying
translation-ergodicity of the coupling distribution.  
In Section~\ref{sec:standard} we presented an explicit construction
of such a non-self-averaged thermodynamic state
$\rho_{\cal J}$, which obeyed
the physically important requirement of translation-covariance,
and whose overlap distribution function was 
thus translation-invariant.
We know of no other (natural) construction of a thermodynamic state
for the EA model (which is measurable with
respect to the disorder configuration), in the event that the
spin glass does indeed possess many states (in which case 
chaotic size dependence must be taken into account)
at some dimension and temperature.

Although we presented these results (and much of
our other discussion on spin glasses) in the
context of the EA model, we stress that they
apply quite generally to most realistic
spin glass models, because they depend only
on general properties such as translation-invariance
of the overlap function and translation-ergodicity
of the underlying disorder distribution.

These results lead, however, to a new approach to
the thermodynamics of systems with many competing states that
is far more general than considerations of spin
glasses alone might indicate.  The failure of
the standard SK picture arises from the fact
that if many pure states do exist for a particular
system (at some dimension and temperature), then
chaotic size dependence generally follows and it
becomes unreasonable to describe the thermodynamics
through a single state --- even though this state
may be a mixture of infinitely many pure states ---
as in the standard approach.  As an alternative,
and based on the example of a chaotic dynamical system, we 
describe the thermodynamics through an {\it ensemble
of states\/} (which may themselves be mixtures of pure
states) that we call the metastate.
Within that approach, the idea
of replicas (whose correlations determine
the metacorrelations of the metastate)
becomes natural and formerly mysterious
concepts --- such as replica symmetry breaking --- become clear.
Further, the connections between these and newer concepts
such as replica non-independence and dispersal
of the metastate can be easily uncovered.

A crucial issue is the replacement of the old concept
of non-self-averaging (as dependence on the bulk
coupling realization) with a new version
of dependence on boundary conditions at infinity.
This allows for the possibility that 
moments of $q$ (for example) as computed through
the distribution $P_{\cal J}^{(L)}(q)$ in any finite
volume can depend on ${\cal J}$ for arbitrarily large $L$ ---
even though $P_{\cal J}(q)$ itself is independent of ${\cal J}$.  Within
the context of the nonstandard SK model, we replace
the idea of dependence on ${\cal J}$ with dependence on
the state $\Gamma$ within the metastate for {\it fixed\/} ${\cal J}$.
This new notion corresponds, roughly speaking, to
dependence on couplings at infinity (which yield
a kind of annealed boundary condition at infinity)
or to dependence on $L$, all for fixed ${\cal J}$.

Applying these results to the EA model, we find that
several scenarios for the metastate remain as
logical possibilities in various dimensions and
temperatures.  One of course is the trivial paramagnetic
phase.  Another is the scaling/droplet model.
Two other possibilities, mentioned in Ref.~\cite{NS96c},
involve states $\Gamma$ consisting of a continuum
of pure states; in one of these scenarios the metastate 
is dispersed and in the other it isn't, although both
exhibit replica symmetry breaking.  However, we see no evidence
that either of these apply to realistic spin
glasses, and so do not discuss them further here.

An intriguing new possibility, also discussed
in Ref.~\cite{NS96c}, is the chaotic pairs picture,
which is different from both droplet/scaling
and mean-field pictures.  This picture follows
naturally from our earlier discussion on the
metastate; it has infinitely many pure states,
but with weights so mismatched in any finite volume (with,
say, periodic boundary conditions) that only a pair of
pure states (related by a global spin flip)
appear.  So in finite volumes this picture
resembles droplet/scaling, but it has a
very different thermodynamics; in particular,
there are infinitely many pure state pairs and
which of these appears in a 
given volume depends chaotically on $L$.  It is
interesting to note that this scenario
actually arises in high dimensions in
a highly disordered ground state model
of spin glasses \cite{NS94}.

Finally, we discussed a maximal mean-field
picture called the nonstandard SK picture.
This picture has features which resemble
some of the familiar aspects of  Parisi-type
spin glass ordering in finite volume --- and
is consistent with various numerical
simulations which claim to see this type
of ordering --- but has an unfamiliar
thermodynamic structure and does not
correspond to the usual picture presented in the
literature.  In particular, it does not
possess nontrivial ultrametricity of all of the pure
states corresponding to a fixed coupling
realization ${\cal J}$; indeed, one of our
results is that such ultrametricity
cannot occur in any reasonable spin glass picture.
It also does not possess non-self-averaging
(in the sense of ${\cal J}$-dependence)
of thermodynamic quantities related to the
order parameter, such as the overlap distribution
function.

Nevertheless, the features of non-self-averaging
and ultrametricity could appear in any
{\it finite\/} volume if this picture should hold.
This leads to a further conclusion, namely, that
for systems with quenched disorder (and for
inhomogeneous systems in general) with many
competing thermodynamic states, {\it properties
which persist in large finite volumes cannot
be straightforwardly extrapolated to a
description of the thermodynamics\/}.
In these cases, the metastate approach is indispensible
for sorting out the thermodynamics.

In any case, we have serious reservations about the viability
of the nonstandard SK picture.  Although we cannot
at this point rule it out on purely logical grounds,
it requires an enormous number of constraints to
be simultaneously satisfied.  

Further work is needed to determine which of
these remaining pictures does hold for real
spin glasses.  Work is also needed to
study the connections between the approach
presented in this paper to systems in equilibrium
and the dynamical behavior of systems out of
equilibrium. Such investigations are currently in progress.
We conclude by again pointing out that although
we have concentrated in this paper on spin glasses,
the phenomenon of thermodynamic chaos and the
metastate approach to its analysis are potentially
applicable to any thermodynamic system (disordered
or not, inhomogeneous or not) in which there are
many competing pure states and the finite volume
boundary conditions (or fields) are not (or cannot be)
carefully chosen to favor one (or a very few) of them.

{\it Acknowledgments.\/}  This research was partially 
supported by NSF Grant DMS-95-00868 (CMN) 
and by DOE Grant DE-FG03-93ER25155 (DLS).


\begin{references}

\small

\bibitem{Ryan92}  For a recent review of a variety
of experimental tests, see 
{\em Recent Progress in Random Magnets\/}, D.H.~Ryan, ed. 
(World Scientific, Singapore, 1992).

\bibitem{BY}
K.~Binder and A.P.~Young,
\newblock {\em Rev.~Mod.~Phys.\/} {\bf 58}, 801 (1986).

\bibitem {EA} 
S.~Edwards and P.W.~Anderson, 
\newblock {\em J.~Phys.~F\/} {\bf 5}, 965 (1975).

\bibitem {SK} 
D.~Sherrington and S.~Kirkpatrick, 
\newblock {\em Phys.~Rev.~Lett.\/} {\bf 35}, 1972 (1975).

\bibitem{MEF}
M.E.~Fisher and R.R.P.~Singh, in {\em Disorder in Physical Systems}, 
edited by G.~Grimmett and D.J.A.~Welsh (Clarendon Press, Oxford, 1990), p.~87.

\bibitem{MPR}
E.~Marinari, G.~Parisi, and F.~Ritort, 
\newblock {\em J.~Phys.~A\/} {\bf 27}, 2687 (1994).

\bibitem{TH}
M.J.~Thill and H.J.~Hilhorst,
\newblock {\em J.~Phys.~I} {\bf 6}, 67 (1996).

\bibitem{NS94}
C.M.~Newman and D.L.~Stein, 
\newblock {\em Phys.~Rev.~Lett} {\bf 72}, 2286 (1994). 

\bibitem{NS96a}
C.M.~Newman and D.L.~Stein, 
\newblock{\em J.~Stat.~Phys.} {\bf 82}, 1113 (1996).

\bibitem{Palmer}
R.G.~Palmer,
\newblock {\em Adv.~Phys.\/} {\bf 31}, 669 (1982).

\bibitem{SA}
P.~Sibani and J.-O.~Andersson,
\newblock {\em Physica~A\/} {\bf 206}, 1 (1994).

\bibitem{NS95}
D.L.~Stein and C.M.~Newman, 
\newblock {\em Phys.~Rev.~E\/} {\bf 51}, 5228  (1995). 

\bibitem{Orbach1}
M.~Lederman, R.~Orbach, J.M.~Hamann, M.~Ocio, and E.~Vincent,
\newblock {\em Phys.~Rev.~B\/} {\bf 44}, 7403 (1991).

\bibitem{Orbach2}
Y.G.~Joh, R.~Orbach, and J.M.~Hamann, ``Spin glass dynamics
under a change in magnetic field'', preprint, 1996.

\bibitem{aging1}
P.~Refrigier, E.~Vincent, J.~Hamman, and M.~Ocio, 
\newblock {\em J.~Phys.~(Paris)\/} {\bf 48}, 1533 (1987).

\bibitem{aging2}
G.J.M.~Koper and H.J.~Hilhorst, 
\newblock {\em J.~Phys.~(Paris)\/} {\bf 49}, 249 (1988).

\bibitem{aging3}
P.~Sibani and K.-H.~Hoffmann, 
\newblock {\em Phys.~Rev.~Lett.\/} {\bf 63}, 2853 (1989).

\bibitem{aging4}
K.-H.~Hoffmann and P.~Sibani, 
\newblock {\em Z. Phys. B} {\bf 80}, 429 (1990).

\bibitem{aging5}
P.~Svedlinh, K.~Gunnarson, J.-O.~Andersson, H.A.~Katori, and A.~Ito, 
\newblock {\em Phys.~Rev.~B\/} {\bf 46}, 13687 (1992).

\bibitem{aging6} 
J.P.~Bouchaud, 
\newblock {\em J.~Phys.~I} {\bf 2}, 1705 (1992).

\bibitem{aging7}
F.~Lefloch, J.~Hamann, M.~Ocio, and E.~Vincent, 
\newblock {\em Europhysics~Lett.\/} {\bf 18}, 647 (1992).

\bibitem{aging8}
H.~Rieger, 
\newblock {\em J.~Phys.~A} {\bf 26}, L615 (1993).

\bibitem{aging9}
S.~Franz and M.~M\'ezard, 
\newblock {\em Physica A\/} {\bf 210}, 48 (1994).

\bibitem{aging10}
E.~Vincent, J.P.~Bouchaud, D.S.~Dean, and J.~Hamann,
\newblock {\em Phys.~Rev.~B} {\bf 52}, 1050 (1995).

\bibitem{NS96b}  
C.M.~Newman and D.L.~Stein,
\newblock {\em Phys.~Rev.~Lett.\/} {\bf 76}, 515 (1996).

\bibitem{NS96c}  
C.M.~Newman and D.L.~Stein,
\newblock {\em Phys.~Rev.~Lett.\/} {\bf 76}, 4821 (1996).

\bibitem{AW}
M.~Aizenman and J.~Wehr,
\newblock {\em Comm.~Math.~Phys.\/} {\bf 130}, 489 (1990).
 
\bibitem{NS92}  
C.M.~Newman and D.L.~Stein, 
\newblock {\em Phys.~Rev.~B} {\bf 46}, 973 (1992). 

\bibitem{MPV}
M.~M\'ezard, G.~Parisi, and M.A.~Virasoro,
{\em Spin Glass Theory and Beyond\/} (World Scientific, Singapore, 1987).

\bibitem {Parisi1} 
G.~Parisi, 
\newblock {\em Phys.~Rev.~Lett.\/} {\bf 43}, 1754 (1979). 

\bibitem{Parisi2}
G.~Parisi, 
\newblock {\em Phys.~Rev.~Lett.\/} {\bf 50}, 1946 (1983).

\bibitem{Houghton83}
A.~Houghton, S.~Jain, and A.P.~Young, 
\newblock {\em J.~Phys.~C\/} {\bf 16}, L375 (1983).

\bibitem {MPSTV}
M.~M\'ezard, G.~Parisi, N.~Sourlas, G.~Toulouse, and M.~Virasoro, 
\newblock {\em Phys.~Rev.~Lett.\/} {\bf 52}, 1156 (1984). 

\bibitem{Mac} 
W.L.~McMillan, 
\newblock {\em J.~Phys.~C\/} {\bf 17}, 3179 (1984).

\bibitem {BM}
A.J.~Bray and M.A.~Moore,
\newblock {\em  Phys.~Rev.~Lett.\/} {\bf 58}, 57 (1987).

\bibitem {FH86} 
D.S.~Fisher and D.A.~Huse, 
\newblock {\em Phys.~Rev.~Lett.\/} {\bf 56}, 1601 (1986).

\bibitem {FH88} 
D.S.~Fisher and D.A.~Huse, 
\newblock {\em Phys.~Rev.~B\/} {\bf 38}, 386 (1988).

\bibitem {HF87a} 
D.A.~Huse and D.S.~Fisher, 
\newblock {\em J.~Phys.~A\/} {\bf 20}, L997 (1987).

\bibitem {HF87b}
D.S.~Fisher and D.A.~Huse, 
\newblock {\em J.~Phys.~A\/} {\bf 20}, L1005 (1987).

\bibitem{E1990}
A.~van~Enter,
\newblock {\em J.~Stat.~Phys.\/} {\bf 60}, 275 (1990).

\bibitem{Alm78}
J.R.L.~de~Almeida and D.J.~Thouless,
\newblock {\em J.~Phys.~A} {\bf 11}, 983 (1978).

\bibitem{Caracciolo}
S.~Caracciolo, G.~Parisi, S.~Patarnello, and N.~Sourlas, 
\newblock {\em J.~Phys.~France\/} {\bf 51}, 1877 (1990).

\bibitem{FHcrit}
D.A.~Huse and D.S.~Fisher,
\newblock {\em J.~Phys.~I} {\bf 1}, 621 (1991).

\bibitem{Weissman}
M.~Weissman,
\newblock {\em Rev.~Mod.~Phys.\/} {\bf 65}, 829 (1993).

\bibitem{BF1986}
A.~Bovier and J.~Fr\"ohlich,
\newblock {\em J.~Stat.~Phys.\/} {\bf 44}, 347 (1986).

\bibitem{NoteDLR}
Infinite-volume Gibbs measures $\rho_{{\cal J},\beta}$ can also
be characterized, without such a limiting process, as
probability measures (on infinite-volume spin configurations)
which satisfy the Dobrushin-Lanford-Ruelle (DLR) equations.
For a mathematically detailed presentation, see \cite{Georgii}.

\bibitem{Georgii}
H.O. Georgii, 
\newblock {\em Gibbs Measures and Phase Transitions\/}
(de Gruyter Studies in Mathematics, Berlin, 1988).

\bibitem{vEvH}
A.C.D.~van~Enter and J.L.~van~Hemmen,
\newblock{\em Phys.~Rev.~A} {\bf 29}, 355 (1984).

\bibitem{NSBerlin}
C.M.~Newman and D.L.~Stein, in {\it Mathematics of Spin
Glasses and Neural Networks\/}, edited by A.~Bovier and
P.~Picco (Birkh\"{a}user, Boston, to appear).

\bibitem{Zurich}
C.M.~Newman, 
\newblock {\em Topics in Disordered Systems\/} (Birkh\"auser, 
Basel, to appear).

\bibitem{Wiener1939}
N.~Wiener, 
\newblock {\em Duke Math. J.\/}  {\bf 5}, 1 (1939).

\bibitem{MP}
M.~M\'ezard and G.~Parisi, 
\newblock {\em J.~Phys.~I France\/} {\bf 1}, 809 (1991).

\bibitem{BMY1}
J.-P.~Bouchaud, M.~Mezard, and J.S.~Yedidia,
\newblock {\em Phys.~Rev.~Lett.\/} {\bf 67}, 3840 (1991).

\bibitem{BMY2}
J.S.~Yedidia, in {\it 1992 Lectures in Complex Systems\/},
edited by D.L.~Stein
(Addison-Wesley, Reading, MA, 1993), p.~299.

\bibitem{Parisi3}
G.~Parisi,
\newblock{\em Physica A\/} {\bf 194}, 28 (1993).

\bibitem{VHO} 
E.~Vincent, J.~Hammann, and M.~Ocio, in Ref.~\cite{Ryan92}, p.~207.

\bibitem{BCPR}
D.~Badoni, J.C.~Ciria, G.~Parisi, F.~Ritort, J.~Pech, and J.J.~Ruiz-Lorenzo,
\newblock {\em Europhys.~Lett.\/} {\bf 21}, 495 (1993).

\bibitem{FPV}
S.~Franz, G.~Parisi, and M.A.~Virasoro,
\newblock {\em J.~Phys.~I France\/} {\bf 4}, 1657 (1994).

\bibitem{Ritort94}
F.~Ritort, 
\newblock {\em Phys.~Rev.~B\/} {\bf 50}, 6844 (1994).

\bibitem {LD}
P.~Le~Doussal and T.~Giamarchi,
\newblock {\em Phys.~Rev.~Lett.\/} {\bf 74}, 606 (1995).

\bibitem{MPRR}
E.~Marinari, G.~Parisi, J.J.~Ruiz-Lorenzo, and F.~Ritort,
\newblock {\em Phys.~Rev.~Lett.\/} {\bf 76}, 843 (1996).

\bibitem{Brout}
R.~Brout,
\newblock {\em Phys.~Rev.\/} {\bf 115}, 824 (1959).

\bibitem{Kac}
M.~Kac, {\em Trondheim Theoretical Physics Seminar\/}
(Nordita, Publ.~No.~286, 1968).

\bibitem{Edwards}
S.F.~Edwards, in {\em Proceedings of the Third International
Conference on Amorphous Materials\/}, edited by R.W.~Douglas
and B.~Ellis (Wiley, NY, 1970), p.~279; also in {\em Polymer Networks\/},
edited by A.J.~Chompff and S.~Newman (Plenum, NY, 1971), p. 83.

\bibitem{Blandin}
A.~Blandin,
\newblock {\em J.~Phys.~(Paris)~Colloq.\/} {\bf C6-39}, 1499 (1978).

\bibitem{DT}
B.~Derrida and G.~Toulouse,
\newblock {\em J.~Phys.~(Paris)~Lett.\/} {\bf 46}, L223 (1985).

\bibitem{MPV86}
M.~M\'ezard, G.~Parisi, and M.A.~Virasoro,
\newblock {\em Europhys.~Lett.\/} {\bf 1}, 77 (1986).

\bibitem{TAP}
D.J.~Thouless, P.W.~Anderson, and R.G.~Palmer,
\newblock {\em Phil.~Mag.\/} {\bf 35}, 593 (1977).

\bibitem{BM1980}
A.J.~Bray and M.A.~Moore,
\newblock {\em J.~Phys.~C\/} {\bf 13}, L469 (1980).

\bibitem{BM1981}
A.J.~Bray and M.A.~Moore,
\newblock {\em J.~Phys.~A\/} {\bf 14}, L377 (1981).

\bibitem{MY1982}
N.D.~Mackenzie and A.P.~Young,
\newblock {\em Phys.~Rev.~Lett.\/} {\bf 49}, 301 (1982).

\bibitem{RBY}  
J.D.~Reger, R.N.~Bhatt, and A.P.~Young, 
\newblock {\em Phys.~Rev.~Lett.\/} {\bf 64}, 1859 (1990). 

\bibitem{A1980}
M. Aizenman, 
\newblock {\em Commun.~Math.~Phys.\/} {\bf 73}, 83 (1980).

\bibitem{H1981}
Y. Higuchi,  
in {\em Random Fields, Vol. I\/}, edited by J.~Fritz, J.L.~Lebowitz and
D.~Sz\'asz (North Holland, Amsterdam, 1979), p. 517. 

\bibitem {GKN} 
A.~Gandolfi, M.~Keane, and C.M.~Newman,
\newblock {\em Prob.~Theory~Rel.~Fields\/} {\bf 92}, 511 (1992).

\bibitem{Le1977}
F. Ledrappier, 
\newblock {\em Commun.~Math.~Phys.\/} {\bf 56}, 297 (1977).

\bibitem{Co1989}
F. Comets,
\newblock {\em Prob.~Theory~Rel.~Fields\/} {\bf 80}, 407 (1989).

\bibitem{S1995}
T. Sepp\"{a}l\"{a}inen, 
\newblock {\em Commun.~Math.~Phys.\/} {\bf 171}, 233 (1995).

\bibitem{Ku96}
C.~K\"ulske, in {\it Mathematics of Spin
Glasses and Neural Networks\/}, edited by A.~Bovier and
P.~Picco (Birkh\"{a}user, Boston, to appear).

\bibitem{CMB94}
M.~Cieplak, A.~Maritan, and J.R.~Banavar, 
\newblock {\em  Phys.~Rev.~Lett.\/} {\bf 72}, 2320 (1994).

\bibitem{Guerra}
F.~Guerra, private communication.

\bibitem{Noteoverlaps}  
It is important to mention that the actual overlap
distribution may be sensitive to the limit procedure used in
constructing it.  For example, two different constructions (each of which
yields a well-defined limit for the overlap distribution as $L\to\infty$) were
proposed in Ref.~\cite{NS96b}.  
In these constructions, there are boxes, of sizes $L_a$, $L_b$ and
$L_c$, with periodic b.c.'s imposed on the $L_c$-box, fixed couplings
in the $L_b$-box and overlaps computed in the $L_a$-box.  The first
construction takes $1 \ll L_b \ll L_a = L_c$ while the second construction,
which is the one described in Section IV of this paper, takes
$1 \ll L_a \ll L_b \ll L_c$.
These two constructions are believed to yield different
limit overlap distributions for the random-field
Ising model (A.~van~Enter, private communication) and 
for the highly disordered spin glass model
of Ref.~\cite{NS94} in high dimensions.  Furthermore, they 
possibly would yield different distributions
also for the usual finite-dimensional spin glasses if something like the 
nonstandard SK picture, described in Section~\ref{sec:nonstandard}, holds.
Additional remarks on this issue can be found in 
Parisi's comment (available as cond-mat preprint 9603101 at
http://www.sissa.it/) on Ref.~\cite{NS96b} and in Newman and Stein's
reply (available as adap-org preprint 9603001
at http://xxx.lanl.gov/).

\bibitem{PS}
This was proved for the SK model
by L.~Pastur and M.~Shcherbina, {\it J.~Stat.~Phys.\/}
{\bf 62}, 1 (1991).

\bibitem{Notediscrete}
We note, though, that what
is actually required is discreteness of the {\it overlap
distribution\/}, and it has been pointed out \cite{EnHM1992} that this could be the case
without discreteness of the pure state decomposition
Eq.~(\ref{eq:sumgamma}).  
However, there is another feature of the standard SK picture
which seems to require discreteness of at least the low-lying
part of the energy (or free energy) spectrum of pure states.
This is the occurrence of energy (or free
energy) gaps of order unity between the low-lying states
in any (large) volume,
accompanied by an exponentially increasing density
of states near the bottom of the spectrum \cite{DT,MPV86}.
For the remainder of this paper, we will assume 
a countable pure state
decomposition.

\bibitem{EnHM1992}
A.C.D.~van~Enter, A.~Hof and J.~Mi\c{e}kisz, 
\newblock {\em J.~Phys.~A\/} {\bf 25}, L1133 (1992).



\end{references}
\end{document}